\documentclass[seceq]{ptptex}
\newcommand{\siml}{\hspace{.3em}\raisebox{.4ex}{$<$}\hspace{-.75em}
 \raisebox{-.7ex}{$\sim$}\hspace{.3em}}
\newcommand{\simr}{\hspace{.3em}\raisebox{.4ex}{$>$}\hspace{-.75em}
 \raisebox{-.7ex}{$\sim$}\hspace{.3em}}

\usepackage{graphicx}
\usepackage{bm}
\usepackage{ulem}
\usepackage{wrapft}

\title{
The Sign Problem  and MEM in Lattice Field Theory with the $\theta$ Term
}

\author{
 Masahiro Imachi$^{\ddagger}$\footnote{E-mail: imachi@sci.kj.yamagata-u.ac.jp},\\
  Yasuhiko Shinno$^{\diamond}$\footnote{E-mail: 
  shinno@dirac.phys.saga-u.ac.jp} and 
   Hiroshi Yoneyama$^{\diamond}$\footnote{E-mail: yoneyama@cc.saga-u.ac.jp}
}

\inst{
 Department of Physics, Yamagata University, Yamagata 990-8560, Japan$^{\ddagger}$\\
 Department of Physics, Saga University, Saga 840-8502, Japan$^{\diamond}$\\}

\abst{
 Lattice field theory with the  $\theta$ term suffers from the sign
 problem. The sign problem appears as flattening of the free energy. 
 As an alternative to the conventional method, the
 Fourier transform method (FTM), we apply the maximum entropy method
 (MEM) to Monte Carlo data obtained using the CP$^3$ model with the $\theta$
 term. 
 For  data without flattening, we obtain the most probable images of
 the partition function ${\hat{\cal Z}}(\theta)$ with rather small
 errors.  The results are quantitatively close to  the result obtained with the 
 FTM. Motivated by this fact,  we systematically investigate flattening in
 terms of the MEM. 
 Obtained images ${\hat{\cal Z}}(\theta)$ are  consistent with the 
   FTM for  small values of $\theta$, while the behavior of ${\hat{\cal Z}}(\theta)$ 
  depends  strongly on the default model 
 for   large values of $\theta$. This   behavior of ${\hat{\cal Z}}(\theta)$ 
 reflects  the flattening  phenomenon. 
}

\begin{document}

\maketitle

\section{Introduction}
\label{sec:Intro}
 In QCD and the CP$^{N-1}$ model, topologically
 non-trivial configurations play important roles in  determining dynamical properties 
 and the vacuum structure. The effect of these configurations is 
 introduced into the action with a $\theta$ term. The existence
 of the $\theta$ term is associated with several interesting topics,  
 such as the strong CP problem and possible rich phase structures in
 $\theta$ space. It was pointed out by  't Hooft~\cite{rf:tHooft} that
 a color magnetic monopole becomes  a dyon-like object in regions for which $\theta\neq 0$, 
  and this could result in the appearance  of new phase structure. It was
 also shown in Ref. \citen{rf:CR} that a new phase could emerge in the
 Z(N) model. In the CP$^{N-1}$ model, it is known that
 there is a first order phase transition point at $\theta=\pi$ in the
 strong coupling region. To obtain  a  comprehensive understanding of the
 phase structure in $\theta$ space, it is necessary to analyze the phase
 structure in the weak coupling region. 
\par 
 A numerical simulation based upon the importance sampling method is one
 of the most promising tools to study non-perturbative properties of
 field theories. This method, however, is confronted with difficulties
 in the case of theories possessing the $\theta$ term, because this 
 term causes  the Boltzmann weight to be  complex.  
 This is the  ``complex action problem" or the  ``sign problem". A conventional
 way to circumvent this problem is to calculate the partition function
 ${\cal Z}(\theta)$ by Fourier-transforming the topological charge
 distribution $P(Q)$, which is calculated with the real positive
 Boltzmann weight.\cite{rf:BRSW,rf:Wiese,rf:HITY,rf:PS,rf:FO}
 We call this  the ``Fourier transform method'' (FTM). 
 Although this approach works well for small lattice volumes and/or in the
 strong coupling region, it does not work for large volumes and/or in
 the weak coupling region,  due to  flattening of the free energy
 density $f(\theta)$. This  flattening phenomenon results from the error
 in $P(Q)$ obtained using Monte Carlo (MC) simulations and 
  leads to a spurious  phase transition for $\theta=\theta_{\rm
 f} (<\pi)$. This is the sign problem. 
 To overcome this problem requires exponentially increasing statistics.
\par   
 As an alternative approach to the FTM,\cite{rf:ACGL,rf:AANV} we have
 applied the maximum entropy method (MEM) to the treatment of the sign problem.   
  This method  is based upon Bayes' theorem, and  it  has 
 been widely used in various 
 fields.\cite{rf:Bryan,rf:JG,rf:AHN,rf:Yamazaki_CP-PACS,rf:Fiebig,rf:ACHKS,rf:IS,rf:MDJM,rf:SSH,rf:EH,rf:UM,rf:ISY}  
 The MEM  gives the most probable images  
 by utilizing sets of data and our knowledge about these images. The
 probability distribution function, called the posterior 
 probability, is given by the product of the likelihood function and the
 prior probability. The latter represents our knowledge about the
 image and the former indicates how data points are  distributed  around
 the true values. The prior probability is
 given as an entropy term, which plays the essential role to guarantee
 the uniqueness of the solution. 
\par
 In order to investigate whether the MEM is applicable to the sign problem, we
 applied it  to mock data, which were  prepared by adding Gaussian noise
 to a model. In Ref. \citen{rf:ISY}, a Gaussian $P(Q)$ was used. 
 The corresponding  free energy can be calculated analytically  using the 
 Poisson sum formula. As mock data, data with flattening and without
 flattening were prepared. We found that in both  cases, the MEM reproduced smooth
 $f(\theta)$. 
 The values of obtained  $f(\theta)$
  are  in agreement with exact ones  and the 
 errors are reasonably small compared to those  resulting when using the Fourier
 transform. These results might lead  one to believe that  the MEM has the effect of 
  smoothing  data and that, for this reason,  it  is not a suitable technique for
 detecting singular behaviors,  such as  phase transitions. 
 To determine whether this is indeed the case,  we analyzed some toy models that 
  exhibit singular behavior originating  from the  characteristics of the models 
 themselves.\cite{rf:ISY2} We found in Ref. \citen{rf:ISY2} that  in fact,   the MEM
 can detect such singular behavior. 
\par
 In the present paper, we apply the MEM to MC data of the CP$^3$
 model. 
 In order  to check whether the MEM can treat  real data in the
 $\theta$ term,  data without flattening are used. Next, we
 investigate how the flattening phenomenon is observed within  the
 MEM. In the MEM analysis, it is necessary to give prior
 information. Generally, an obtained image depends on this  prior
 information. We systematically investigate the influence of this 
 information on the most probable image. The uncertainty in  the most
 probable image is estimated as an error. We also check the
 effectiveness of the MEM by  considering this error. 
\par
 This paper is organized as follows. In the following section, we
 summarize the  formulations used in this work. Numerical results are presented
 in \S\ref{sec:Results}. In that  section, we investigate in detail the behavior of 
 the obtained most probable images.  In the final
 section, conclusions and discussion are presented.
\par
\setcounter{equation}{0}

\section{Formulations}
\label{sec:Formula}
\subsection{Topological charge distribution}
\label{sub:charge}
 The  lattice action of the CP$^{N-1}$ model with the $\theta$ term is
 defined by  
\begin{equation}
 S_{\theta}({\bar z},z)=S({\bar z},z)-i\theta Q({\bar z},z), 
  \label{eqn:thetaaction}
\end{equation}
 where $S$ is a lattice action and $Q$ is a topological charge. Complex
 scalar fields of the model are denoted by ${\bar z}$ and $z$, where ${\bar z}$ is
 the complex conjugate of $z$. \par
 We choose an integer-valued topological charge,\cite{rf:BL} 
\begin{equation}
 Q=\frac{1}{2\pi}\sum_{\Box}A_{\Box}, \label{eqn:topcha}
\end{equation}
 where the plaquette contribution $A_{\Box}$ is
 given by 
\begin{equation}
 A_{\Box}=\frac{1}{2}\sum_{\mu,\nu}\left\{A_{\mu}(n)+A_{\nu}(n+{\hat \mu})
	   -A_{\mu}(n+\nu)-A_{\nu}(n)\right\}. ~~~({\rm mod}~2\pi)
 \label{eqn:geocha}
\end{equation}
 Here we have  $A_{\mu}(n)\equiv \arg[{\bar z}(n)z(n+{\hat \mu})]$, and these quantities 
 satisfy $A_{\mu}(n)\in [-\pi,\pi]$.\par
 As the conventional method  for   avoiding  the complex Boltzmann weight, the
 partition function ${\cal Z}(\theta)$ is calculated by
 Fourier-transforming the topological charge distribution $P(Q)$:
\begin{equation}
 {\cal Z}(\theta)=\sum_Q P(Q)e^{i\theta Q}. \label{eqn:partition}
\end{equation}
 The distribution $P(Q)$ is given by
\begin{equation}
 P(Q)\equiv \frac{\int[d{\bar z}dz]_Q e^{-S({\bar z},z)}}
  {\int[d{\bar z}dz]e^{-S({\bar z},z)}}, \label{eqn:Pq}
\end{equation}
 which is calculated with a real positive Boltzmann weight. 
 The measure $[d{\bar z}dz]_Q$ expresses the meaning  that the integral is
 restricted to configurations of ${\bar z}$ and $z$ with $Q$. Note that
 $P(Q)$ is normalized as $\sum_Q P(Q)=1$.\par
 We update configurations with  the combined use
 of the overrelaxation and  Metropolis algorithms. From the
 generated configurations, we measure $Q$ and construct a histogram by
 counting the number of  configurations with $Q$. 
 Because the  $P(Q)$ under consideration decreases rapidly as a function of $Q$, it is
 convenient to use the set method\cite{rf:KSC}, in which an entire range
 of values of  $Q$ is divided 
 into sets $S_i~(i=1,~2,~3,\cdots)$. In the present study, each of the sets
 $S_i$ consists of 4 bins,  $Q=3i-3,~3i-2,~3i-1,~\mbox{and}~3i$, 
  and thus each set shares its two external bins with the 
  adjacent sets. Explicitly, the shared bins are 
 $Q=3k~(k=1,~2,~3,\cdots)$. 
 In order to generate configurations more effectively and to reduce
 errors, the action is modified by adding a trial function
 $P_t(Q)$ satisfying 
\begin{equation}
 S_{eff}=S-\ln P_t(Q). \label{eqn:effaction}
\end{equation}
 The form of $P_t(Q)$ is chosen as  $P_t(Q)\propto e^{-\alpha Q^2}$ in
 the present study, where $\alpha$ is adjusted so that $P(Q)$ becomes
 almost flat in any given  set,  in order to reduce errors. The power 2 of $Q$ in
 $P_t(Q)$ is varied  in a manner depending on the coupling
 constant.\cite{rf:BISY}\par 
 For the lattice action, we use a fixed point action (FP
 action)\cite{rf:HN} in order to reduce lattice artifacts. Because  our
 simulations require a large number of measurements due
 to the $\theta$ term, we employ 9 coupling
 constants, which are limited to a short range and lie within one
 plaquette. With this action, it was shown in 
 Ref. \citen{rf:BISY} that the lattice artifact is negligible  up to
 somewhat small coupling constant,  which corresponds to a correlation
 length of several units of the lattice spacing. We refer the reader to
 Ref. \citen{rf:BISY} for actual values of the coupling constants.

\subsection{Flattening of the free energy density}
\label{sub:flat}
 The free energy density $f(\theta)$ is calculated by
 Fourier-transforming $P(Q)$ obtained by MC simulation. 
 The quantity $f(\theta)$ is defined as 
\begin{equation}
 f(\theta)=-\frac{1}{V}\ln\sum_QP(Q)e^{i\theta Q}, \label{eqn:free}
\end{equation}
 where $V=L^2$, the square of the lattice size. \par
 The MC data for $P(Q)$ consist of the true value, ${\tilde P}(Q)$,
 and its error, $\Delta P(Q)$. When the error  at $Q=0$
  dominates because of the exponential damping of $P(Q)$,
 $f(\theta)$ is closely approximated by 
\begin{equation}
 f(\theta)\simeq -\frac{1}{V}\ln\left[e^{-V{\tilde f}(\theta)}+\Delta P(0)
				 \right],
  \label{eqn:approxfree}
\end{equation}
 where ${\tilde f}(\theta)$ is the true value of $f(\theta)$. 
 Because  ${\tilde f}(\theta)$ is an increasing function of $\theta$,  as 
  in the case of ${\tilde f}(\theta)$   derived from  the Gaussian $P(Q)$
   in the strong coupling region,
 $\Delta P(0)$ dominates for large values of $\theta$. If $|\Delta
 P(0)|\simeq e^{-V{\tilde f}(\theta)}$ at $\theta=\theta_{\rm f}$, then 
 $f(\theta)$ becomes almost flat for $\theta\simr \theta_{\rm f}$. This
 is called ``flattening of the free energy density", and  it has been  misleadingly
 identified as a first order phase transition, 
 because the first derivative of $f(\theta)$  appears to jump  at 
 $\theta=\theta_{\rm f}$. 
 To avoid  this problem,
 we must carry out a very large number of  measurements in the FTM; indeed, 
 the order of  $e^V$ measurements are needed.   
\subsection{MEM formalism}
\label{sub:MEM}
 In this subsection, we briefly explain the MEM in terms of the $\theta$
 term. ( For details, see Ref. \citen{rf:ISY}. )\par
 In a parameter inference,  such as the $\chi^2$ fitting, the inverse
 Fourier transform   
\begin{equation}
 P(Q)=\int_{-\pi}^{\pi}\frac{d\theta}{2\pi}{\cal Z}(\theta)e^{-i\theta Q}
  \label{eqn:invfourier}
\end{equation}
 is used. In the  numerical calculations, we use the discretized version
 of Eq. (\ref{eqn:invfourier}); $P(Q)=\sum_n K_{Q,n}{\cal Z}_n$,
 where $K_{Q,n}$ is the Fourier integral kernel and ${\cal Z}_n\equiv
 {\cal Z}(\theta_n)$. In order for  the continuous function ${\cal Z}(\theta)$
  to be reconstructed, a sufficient number of values  of 
 $\theta$, which we denote  by $N_{\theta}$, is required so that the relation $N_{\theta}>N_Q$ holds,
  where $N_Q$ represents the number of data points in
 $P(Q)$ ($Q=0,~1,\cdots,~N_Q-1$). A straightforward application of the
 $\chi^2$ fitting to the case $N_{\theta}>N_Q$ leads to degenerate
 solutions. 
 This is an ill-posed problem. \par
 The maximum entropy method is one  promising tool to solve  this 
 ill-posed problem,  and it gives a unique solution. 
 The MEM is based upon Bayes' theorem, expressed as 
\begin{equation}
 {\rm prob}({\cal Z}(\theta)|P(Q),I)=
  \frac{{\rm prob}(P(Q)|{\cal Z}(\theta),I)~{\rm prob}({\cal Z}(\theta)|I)}
  {{\rm prob}(P(Q)|I)}, \label{eqn:posterior}
\end{equation}
 where ${\rm prob}(A|B)$ is the conditional probability
 that $A$ occurs under the condition that $B$ occurs. 
 The posterior probability ${\rm prob}({\cal Z}(\theta)|P(Q),I)$ is the
 probability that the partition function ${\cal Z}(\theta)$ is realized  when the 
 MC data $\{P(Q)\}$ and prior information $I$ are given. 
 The likelihood function ${\rm prob}(P(Q)|{\cal Z}(\theta),I)$ is given
 by  
\begin{equation}
 {\rm prob}(P(Q)|{\cal Z}(\theta),I)=\frac{1}{X_L}e^{-\frac{1}{2}\chi^2},
  \label{eqn:likelihood}
\end{equation}
 where $X_L$ is a normalization constant and $\chi^2$ is a standard
 $\chi^2$ function. \par 
 The probability ${\rm prob}({\cal Z}(\theta)|I)$, which guarantees the
 uniqueness of the solution, is given in terms of  an
 entropy $S$ as
\begin{equation}
 {\rm prob}({\cal Z}(\theta)|I)
=\frac{1}{X_s(\alpha)}e^{-\alpha S}, \label{eqn:prior}
\end{equation}
 where $\alpha$ and $X_S(\alpha)$ are a positive parameter and an
 $\alpha$-dependent normalization constant, respectively. As $S$, the
 Shannon-Jaynes entropy is conventionally employed:
\begin{equation}
 S=\sum_{n=1}^{N_{\theta}}\left[{\cal Z}_n-m_n
    -{\cal Z}_n\ln\frac{{\cal Z}_n}{m_n}\right].
 \label{eqn:SJentropy}
\end{equation}
Here  $m_n\equiv m(\theta_n)$ represents  a default model. \par
 The posterior probability ${\rm prob}({\cal Z}_n|P(Q),I)$, thus, is given by 
\begin{equation}
 {\rm prob}({\cal Z}_n|P(Q),I,\alpha,m)=\frac{1}{X_LX_s(\alpha)}
  e^{-\frac{1}{2}\chi^2+\alpha S}\equiv\frac{e^{W[{\cal Z}]}}{X_LX_s(\alpha)},
 \label{eqn:posterior2}
\end{equation}
 where it is explicitly expressed  that $\alpha$ and $m$ are regarded
 as new knowledge in \\ ${\rm prob}({\cal Z}_n|P(Q),I,\alpha,m)$.  
 For the prior information $I$, we impose the criterion
\begin{equation}
 {\cal Z}_n>0,  \label{eqn:criterion}
\end{equation}
 so that ${\rm prob}({\cal Z}_n\leq 0|I,\alpha,m)=0$. \par
  The most probable image of ${\cal Z}_n$, denoted as ${\hat {\cal
  Z}}_n$, is calculated according to the following
  procedures.\cite{rf:AHN,rf:ISY}.
\begin{enumerate}
 \item Maximizing $W[{\cal Z}]$ to obtain the most probable image 
       ${\cal Z}_n^{(\alpha)}$ for a given $\alpha$:
\begin{equation}
 \frac{\delta}{\delta{\cal Z}_n}W\left[{\cal Z}\right]|_
  {{\cal Z}={\cal Z}^{(\alpha)}}=
  \frac{\delta}{\delta{\cal Z}_n}\left(-\frac{1}{2}\chi^2+\alpha S
  \right)|_{{\cal Z}={\cal Z}^{(\alpha)}}=0. \label{eqn:Zalp}
\end{equation}
 \item Averaging ${\cal Z}^{(\alpha)}_n$ to obtain the
       $\alpha$-independent most probable image ${\cal Z}_n$:\\
\begin{eqnarray}
 {\hat{\cal Z}}_n&=&\int d\alpha~{\cal Z}^{(\alpha)}_n~
  {\rm prob}(\alpha|P(Q),I,m).
\label{eqn:finalZ}
\end{eqnarray}
       The range of integration is determined so that  the relation\\
       ${\rm prob}(\alpha|P(Q),I,m)\geq{\rm prob}
       ({\hat \alpha}|P(Q),I,m)/10$ holds, where 
       ${\rm prob}(\alpha|P(Q),I,m)$
       is maximized at $\alpha={\hat \alpha}$. 
 \item Error estimation:\\
       The error of the most probable output image ${\hat {\cal Z}}_n$ is
       calculated as the uncertainty of the image, which takes into
       account the correlations of the images ${\hat {\cal Z}}_n$ among
       various  values of $\theta_n$:
\begin{eqnarray}
 \langle(\delta{\hat{\cal Z}}_n)^2\rangle\equiv\int d\alpha~
  \langle(\delta{\cal Z}_n^{(\alpha)})^2\rangle~{\rm prob}(\alpha|P(Q),I,m).
 \label{eqn:uncertainty}
\end{eqnarray}
 Here  $\delta{\hat{\cal Z}_n}$ and $\delta{\cal Z}_n^{(\alpha)}$
       represent the error in ${\hat{\cal Z}}_n$ and that in 
       ${\cal Z}_n^{(\alpha)}$, respectively.
\end{enumerate}

\subsection{The most probable image and the parameter $\alpha$}
\label{sub:comment}
 In the MEM formalism, a real positive parameter $\alpha$ is  
 introduced. This  parameter  plays the  role of  the  trade-off between
 $S$ and $\chi^2$. The most probable value of $\alpha$ is determined by the
 posterior probability of $\alpha$, ${\rm prob}(\alpha|P(Q),I,m)$, 
 appearing  in Eq. (\ref{eqn:finalZ}). The probability ${\rm
 prob}(\alpha|P(Q),I,m)$ is given by
\begin{eqnarray}
 {\rm prob}(\alpha|P(Q),I,m)\equiv P(\alpha)\propto g(\alpha)
  e^{W(\alpha)+\Lambda(\alpha)}, \label{eqn:posterioralpha}
\end{eqnarray}
 where $W(\alpha)\equiv W[{\cal Z}^{(\alpha)}]$ and
 $2\Lambda(\alpha)\equiv
 \sum_k\ln\{\alpha/(\alpha+\lambda_k(\alpha))\}$. The
 function $\Lambda(\alpha)$ represents the contribution of 
 fluctuations of ${\cal Z}(\theta)$ around ${\cal
 Z}^{(\alpha)}(\theta)$,  and the quantities $\lambda_k(\alpha)$ are the eigenvalues of the
 real symmetric matrix 
\begin{equation}
 \Omega_{n,m}\equiv\frac{1}{2}\sqrt{{\cal Z}_m}\frac{\partial^2\chi^2}
  {\partial{\cal  Z}_m\partial{\cal Z}_n}\sqrt{{\cal Z}_n}
  \vert_{{\cal Z}={\cal Z}^{(\alpha)}}. \label{eqn:Omega}
\end{equation}
 Here, the function $g(\alpha)$
 represents  the prior probability of $\alpha$ and is chosen according to
 prior information. In general, two types of
 $g(\alpha)$ are employed, one according to  Laplace's rule, $g_{\rm
 Lap}(\alpha)={\rm const}$, and one according to Jeffrey's rule, $g_{\rm
 Jef}(\alpha)=1/\alpha$. The latter rule is determined by requiring that
 $P(\alpha)$  be invariant with respect to a change in scale, 
 because $\alpha$ is a scale factor. The former rule  means that we have
 no knowledge about the prior information of $\alpha$. 
 In general, the most probable image ${\hat{\cal Z}}(\theta)$ depends on
 $g(\alpha)$. In the present study, we investigate 
 the sensitivity of  ${\hat{\cal Z}}(\theta)$  to the
 choice of $g(\alpha)$.  This is done by studying the quantity 
\begin{eqnarray}
 \Delta(\theta)\equiv\frac{|{\hat{\cal Z}}_{\rm Lap}(\theta)-
  {\hat{\cal Z}}_{\rm Jef}(\theta)|}
  {({\hat{\cal Z}}_{\rm Lap}(\theta)+{\hat{\cal Z}}_{\rm Jef}(\theta))/2},
  \label{eqn:Delta}
\end{eqnarray}\\
 where ${\hat{\cal Z}}_{\rm Lap}(\theta)$ and ${\hat{\cal Z}}_{\rm
 Jef}(\theta)$ represent  the most probable images  according to  Laplace's rule and 
 Jeffrey's rule, respectively. The quantity $\Delta(\theta)$ is a
 relative difference, which is defined as the absolute value of the
 difference between ${\hat{\cal Z}}_{\rm Lap}(\theta)$ and ${\hat{\cal
 Z}}_{\rm Jef}(\theta)$ divided by the average of the two. 
\setcounter{equation}{0}
\section{Numerical results}
\label{sec:Results}
\subsection{Flattening in the Monte Carlo data}
 In this study, we  carried  out Monte Carlo simulations of the CP$^3$ model with the
 fixed point action. We  fixed the coupling
 constant $\beta$ to 3.0 and the lattice size $L$ to  38 and
 50. The corresponding  correlation length is approximately  7 in units of
 the lattice spacing.  We  employed the set method  
 and the trial function method. The total number of
 measurements for  each set  was  at least on   the order of $10^6$. 
 Parameter values  used in the  simulations are listed in Table \ref{table:parCP3FP},
 where $Q_{\rm min}-Q_{\rm max}$ represents the range of the topological
 charge with  which MC simulations were  performed.
 It is noted that all
 data,  except for $L=38$ and $50$,  were obtained in a  previous
 study\cite{rf:BISY}.\par  
\catcode`#=\active \def#{\phantom{0}}
\begin{table}[ht]
\caption{Parameter values used in the  MC simulations of the CP$^3$ model with the FP
 action. For the MEM analysis, new MC simulations were   
 performed for $L=38$ and 50. }
\vspace*{1mm}
\begin{center}
\begin{tabular}{cccc}
\hline
\hline
$\beta$ & \multicolumn{3}{c}{
 $L$~ : $Q_{\rm min}$--$Q_{\rm max}$:
 total number of measurements~(M/set)} \\
\hline
3.0 & 12 : &0--30~~~~~~: &10.0 \\
    & 24 : &0--18~~~~~~: &10.0 \\
    & 32 : &0--24~~~~~~: &#3.0 \\
    & 38 : &0--27~~~~~~: &#5.0 \\
    & 46 : &0--33~~~~~~: &#1.0 \\
    & 50 : &0--15~~~~~~: &30.0 \\
    & 56 : &0--18~~~~~~: &#5.0 \\
\hline
\end{tabular}
\label{table:parCP3FP}
\end{center}
\end{table}\par
\begin{figure}[h]\hspace*{-10mm}
\hfill
 \begin{minipage}[h!]{.50\hsize}
  \begin{center}\hspace*{-10mm}
   \includegraphics[height=60mm]{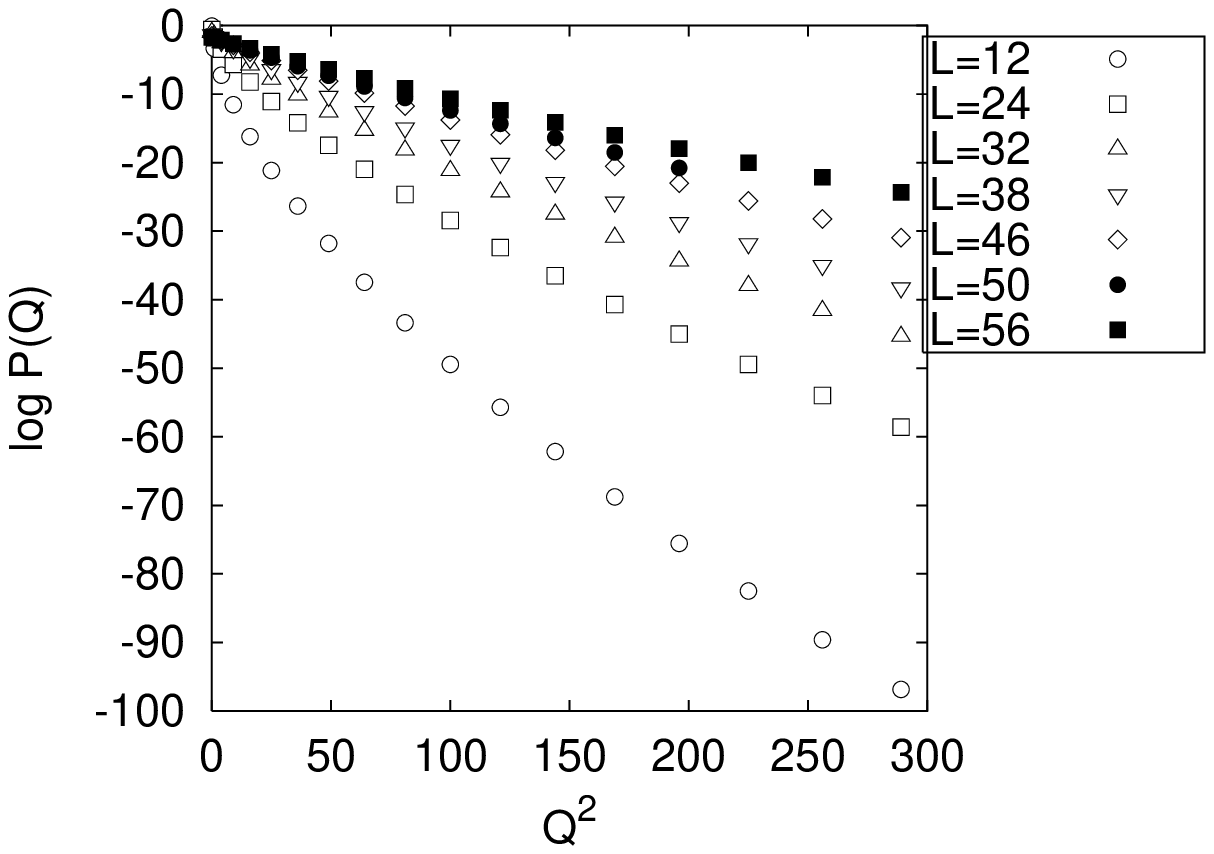}
  \end{center}
 \end{minipage}
 \begin{minipage}[h!]{.50\hsize}
  \begin{center}\hspace*{3mm}
   \includegraphics[height=60mm]{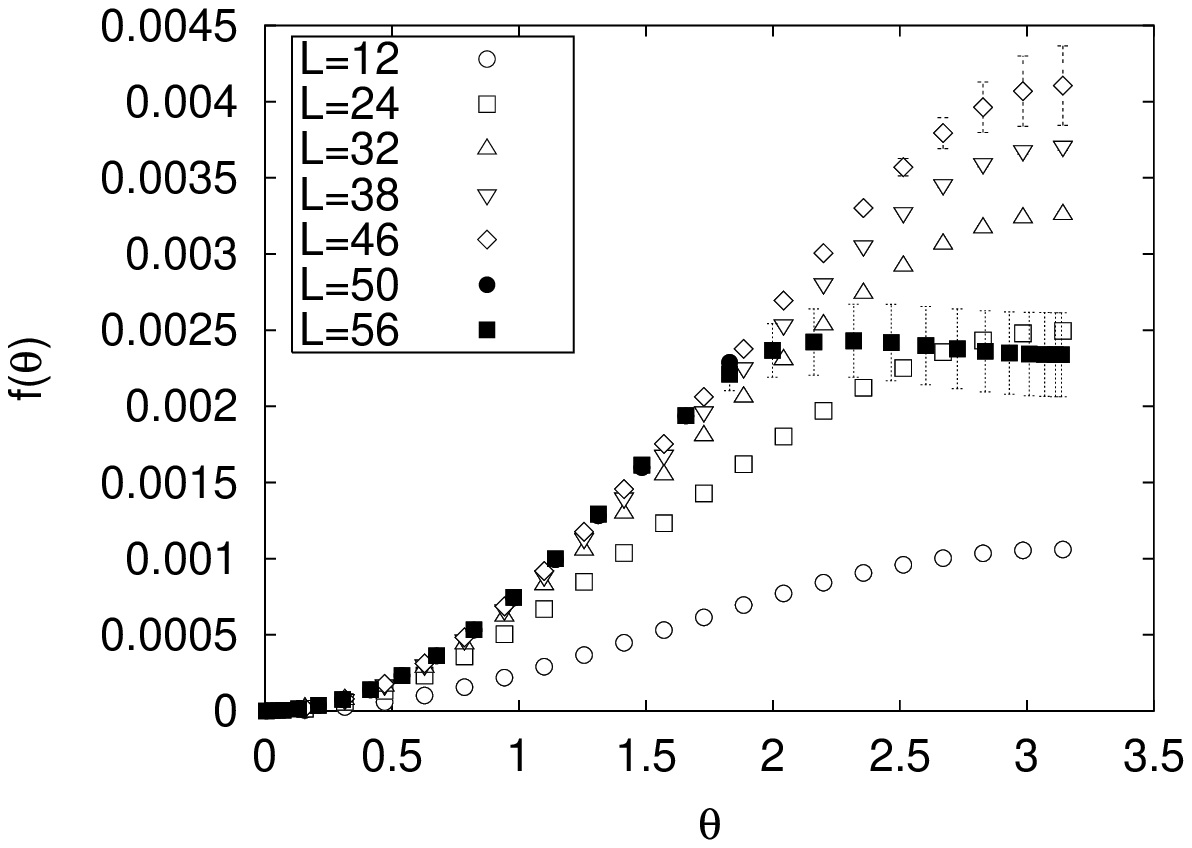}
  \end{center}
 \end{minipage}
 \hfill\vspace*{-1mm}
\caption{Topological charge distributions $P(Q)$ (left panel) and  free
 energy densities $f(\theta)$ (right panel) of the CP$^{3}$
 model for $\beta=3.0$ and various lattice sizes $L$. }
\label{fig:CP3FP}
\end{figure}

 Figure \ref{fig:CP3FP} displays the topological charge distribution
 $P(Q)$ (left panel) and
 the free energy density $f(\theta)$ (right panel) calculated
 by use of the FTM. The error in  $f(\theta)$ was  calculated using  the
 jackknife method.  \par 
 In the right panel of Fig. \ref{fig:CP3FP}, it is observed that
 $f(\theta)$ depends on $L$. The free energy $f(\theta)$
 increases as a function of  $L$. For $\theta\siml 2.0$, $f(\theta)$
 seems to approach an asymptotic line as the  lattice volume increases  
 ($L\geq 38$). For $\theta>2.0$, by contrast, the finite size effect in
 $f(\theta)$ is clearly observed. For $L=56$, flattening is clearly
 seen. For $L=50$, $f(\theta)$ is not plotted for $\theta\simr 2.0$, because  in this case  
   negative values of ${\cal Z}(\theta)$ appear. We also call this behavior
  ``flattening"  for  the same reason that the error in
 $P(Q)$ causes  the FTM to become invalid (see \S\ref{sub:flat}). 
 Although the total number of measurements carried out here  is as large as $3 \times 10^{7}$  in each
 set for $L=50$, flattening is still observed.
\subsection{MEM analysis of the Monte Carlo data}
 As shown in the previous subsection, $f(\theta)$ exhibits flattening
 phenomena for $L=50$ and 56, while  it 
 behaves smoothly for smaller volumes in the FTM. 
 In the present study, we systematically study flattening in terms of
 the MEM. For this purpose, the data for $L=38$ and 50 are used. 
\par
 In our analysis, three types of the default models are used:  (i) $m_{\rm
 c}(\theta)={\rm const}$; (ii) 
 $m_{\rm G}(\theta)=\exp\left[-\gamma\frac{\ln
 10}{\pi^2}\theta^2\right]$;  (iii)
 $m(\theta)={\hat{\cal Z}}(\theta)$ 
 for smaller volumes. In case (i), three values of $m_{\rm c}(\theta)$,
 $1.0,~1.0\times 10^{-3}$ and $1.0\times 10^{-5}$, are employed,  and only
 the results for 
 $m_{\rm c}(\theta)=1.0$ are presented. Case (ii) is the Gaussian
 default model,  and  the parameter
 $\gamma$ in $m_{\rm G}(\theta)$ is varied over a wide range in the
 analysis. In case (iii), to analyze the data for the lattice size $L_0$,
 images  ${\hat{\cal Z}}(\theta)$ obtained from  the MEM analysis for smaller
 volumes are used as default models.  This is because we believe 
 that ${\cal Z}(\theta)$ for smaller 
 volumes might have  properties  similar to those for $L_0$. Such 
 ${\hat{\cal Z}}(\theta)$ may be regarded as prior information. For $L_0=50$,  
 ${\hat{\cal Z}}(\theta)$ for $L=24,~32$ and $38$ are employed as the
 default models. These are denoted as $m_{L/L_0}(\theta)=m_{L/50}(\theta)$. 
  Throughout this  
 paper, it is understood, unless otherwise stated, that  Laplace's rule is used for 
 $g(\alpha)$. The number  $N_Q$ is so chosen that $P(Q)\geq 10^{-18}$
 holds for $L=38$ and $P(Q)\geq 10^{-11}$ holds  for $L=50$. The  
 function $\chi^2$ in Eq. (\ref{eqn:likelihood}) is given in
 terms of the inverse covariance
 matrix of the MC data $\{P(Q)\}$. The inverse matrix is calculated with such 
 precision that the product of the covariance matrix and its
 inverse  has off-diagonal elements that are no larger than  ${\cal
 O}(10^{-30})$. With  these conditions, the value of $N_Q$ is 5 for
 $L=38$ and 7 for $L=50$. 
 It is noted
 that the analysis was  performed with quadruple precision in order  to
 properly reproduce ${\hat{\cal Z}}(\theta)$,  which ranges over many
 orders. 
\subsubsection{Non-flattening case}
 Firstly, we apply the MEM to the data without
 flattening ($L=38$). 
\begin{figure}[h]
\hfill 
 \begin{minipage}[h]{.50\textwidth}
  \begin{center}\hspace*{-15mm}
   \includegraphics[height=60mm]{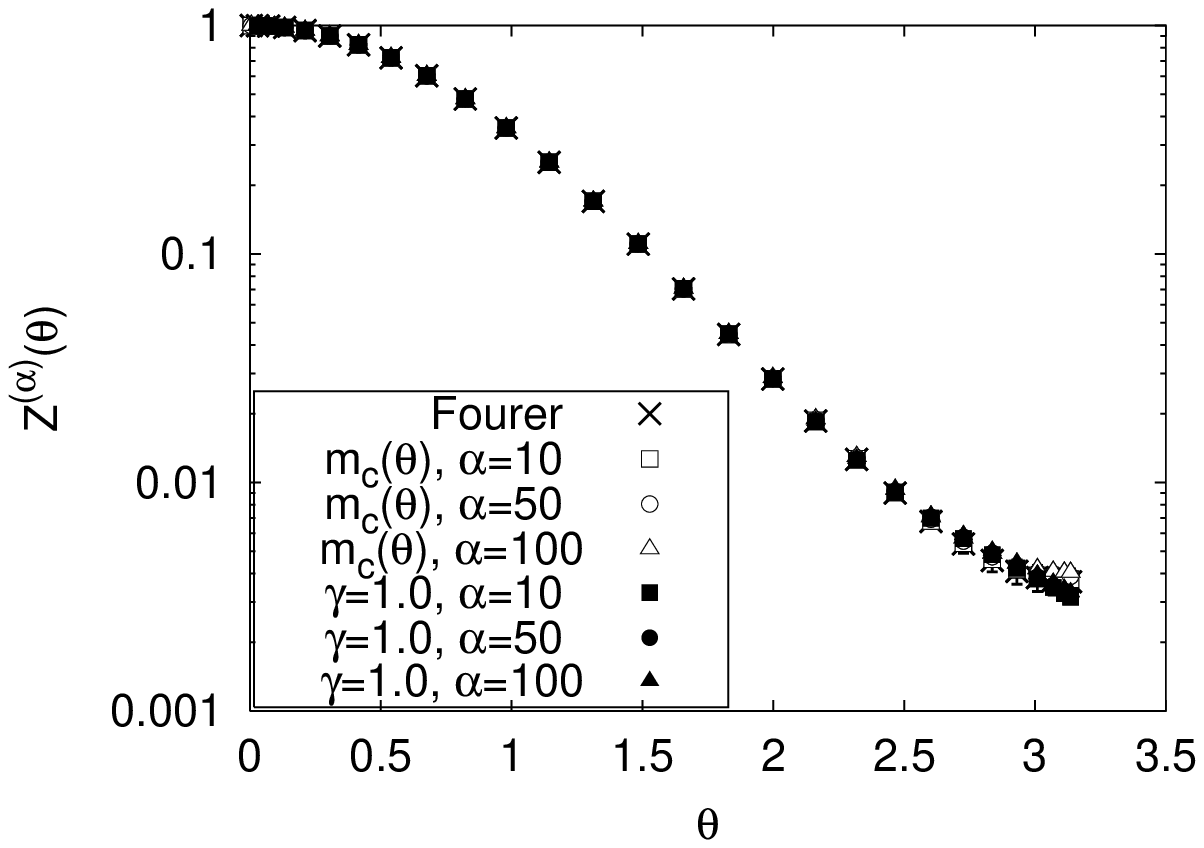}
  \end{center}
 \end{minipage}
 \begin{minipage}[h]{.45\textwidth}
  \begin{center}\hspace*{-5mm}
   \includegraphics[height=60mm]{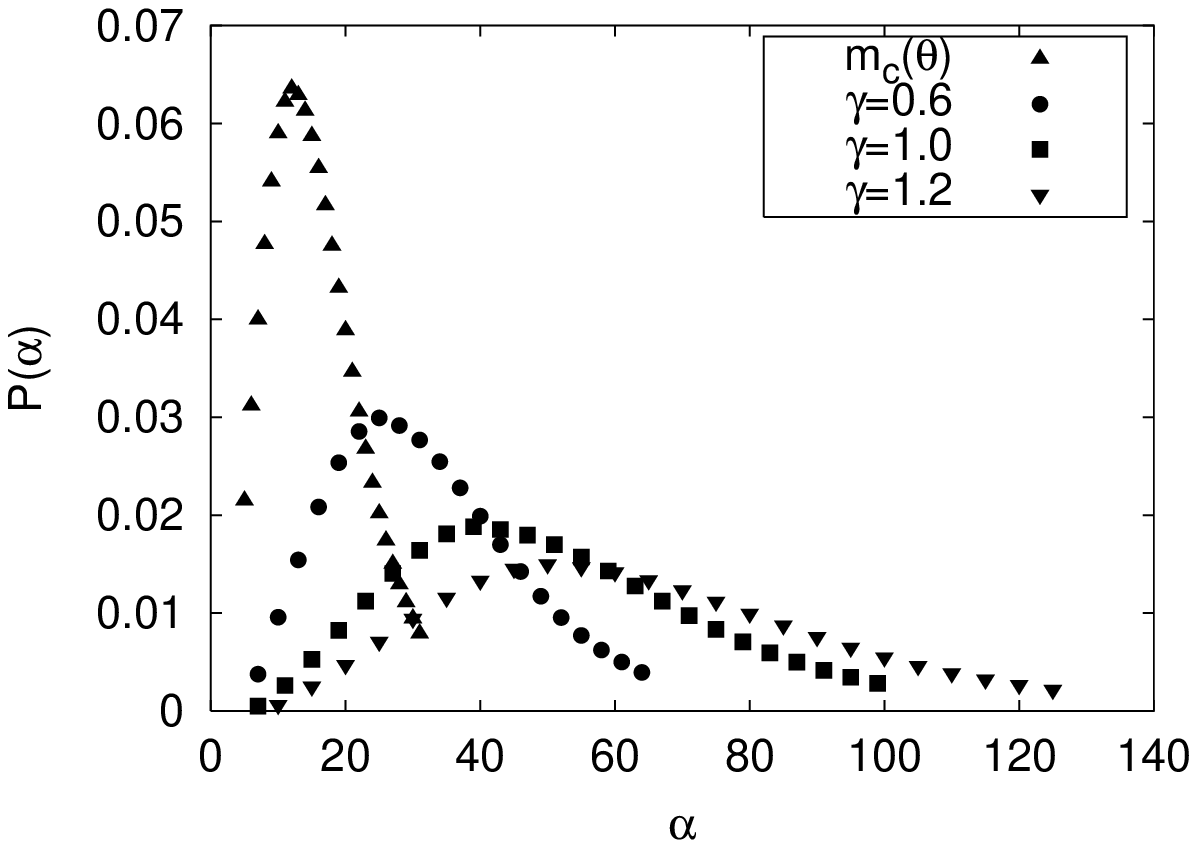}
  \end{center}
 \end{minipage}
\hfill
\caption{Images ${\cal Z}^{(\alpha)}(\theta)$ for a
 given $\alpha$ (left panel) and probabilities $P(\alpha)$
 in the non-flattening case ($L=38$). In the left panel, $m_{\rm c}(\theta)$ and
 $m_{\rm G}(\theta)$ with $\gamma=1.0$ are used. In addition,
 $m_{\rm G}(\theta)$ with $\gamma=0.6$ and 1.2 are used in the right
 panel.} 
\label{fig:Zalpha_L38}
\end{figure}
 The left panel of Fig. \ref{fig:Zalpha_L38} displays ${\cal
 Z}^{(\alpha)}(\theta)$ for a given $\alpha$. 
 Here, $m_{\rm c}(\theta)$ and $m_{\rm
 G}(\theta)$ with $\gamma=1.0$ are used as the default models. The
 partition function obtained using  the FTM, ${\cal Z}_{\rm
 Four}(\theta)$, is also plotted for   comparison. For each $m(\theta)$, the 
 $\alpha$ dependence of ${\cal Z}^{(\alpha)}(\theta)$ is almost
 indiscernible  for $\alpha\in[10,100]$.  It is seen that all the images 
  have only very weak dependence   on   $m(\theta)$ over the entire range of 
 values  of $\theta$. It is also observed that all the results
 of  the MEM are consistent with those  of the FTM. 
 In the right panel, it is seen that $P(\alpha)$ depends on
 $m(\theta)$. The peaks of all the $P(\alpha)$ are located in the range
 $\alpha\leq 50$. The most probable  
 image ${\hat{\cal Z}}(\theta)$ is calculated by using
 Eq. (\ref{eqn:finalZ}). 
 The integrals over  $\alpha$ are trivial,  
 because ${\cal Z}^{(\alpha)}(\theta)$ depends on $\alpha$ only very weakly  over  the
 range of integration. 
 Thus, it is expected that the values of  ${\hat{\cal Z}}(\theta)$ for various
 $m(\theta)$ agree with those of ${\cal Z}_{\rm Four}(\theta)$ over   the entire range of 
 values  of 
 $\theta$. In fact, we find that all the ${\hat{\cal Z}}(\theta)$ are equal to 
  ${\cal Z}_{\rm Four}(\theta)$ within numerical errors. 
 The errors are calculated according to the
 procedure 3 outlined in \S\ref{sub:MEM}. Some of these results are listed in Table
 \ref{table:Z_L38}, specifically,  the values of ${\hat{\cal Z}}(\theta)$ for various
 $m_{\rm G}(\theta)$ and ${\cal Z}_{\rm Four}(\theta)$ at three values
 of $\theta$. 
 These values of $\theta$ are chosen as representatives; the first value is 
 $\theta=2.00$. Up to this value, the asymptotic line of $f(\theta)$ is
 observed (see Fig. \ref{fig:CP3FP}). The second one is $\theta=3.14$ chosen 
 as a value near $\pi$. The third one is $\theta=2.60$, chosen  as a value  approximately 
 halfway  between these two.  It is noted that the errors in 
 ${\hat{\cal Z}}(\theta)$ are rather small for the entire range of values of  $\theta$. 
\par
\catcode`#=\active \def#{\phantom{0}}
\begin{table}[ht]
\caption{Values of the most probable images ${\hat{\cal Z}}(\theta)$ 
 at $\theta=2.00,~2.60~\mbox{and}~3.14$. As default models, 
 $m_{\rm G}(\theta)$ with $\gamma=0.6,~1.0~\mbox{and}~1.2$ were used. For 
   comparison, ${\cal Z}_{\rm Four}(\theta)$ is also listed.}
\begin{center}
\begin{tabular}{c|cccc}
\hline
\hline
$\theta$ & ${\cal Z}_{\rm Four}(\theta)$ & 
 ${\hat{\cal Z}}_{\gamma=0.6}(\theta)$   & 
 ${\hat{\cal Z}}_{\gamma=1.0}(\theta)$   
 & ${\hat{\cal Z}}_{\gamma=1.2}(\theta)$ \\
\hline
 2.00 & 2.840(46)$\times 10^{-2}$ & 2.846(100)$\times 10^{-2}$ 
      & 2.844(81)$\times 10^{-2}$ & 2.844(73)$\times 10^{-2}$ \\
 2.60 & 0.675(47)$\times 10^{-2}$ & 0.696(57)$\times 10^{-2}$ 
      & 0.707(46)$\times 10^{-2}$ & 0.712(41)$\times 10^{-2}$ \\
 3.14 & 0.367(51)$\times 10^{-2}$ & 0.343(52)$\times 10^{-2}$ 
      & 0.322(42)$\times 10^{-2}$ & 0.312(37)$\times 10^{-2}$ \\
\hline
\end{tabular}
\end{center}
\label{table:Z_L38}
\end{table}\par
 The  calculations discussed above were  performed using Eq. (\ref{eqn:finalZ})
 with Laplace's rule. When Jeffrey's rule is employed, the peak of $P(\alpha)$ appears 
 at a value of $\alpha$ smaller than that in the case of  Laplace's rule. The 
 probability $P_{\rm Lap}(\alpha)$ for $\gamma=1.2$, for example, peaks at
 $\alpha=50$, while  $P_{\rm Jef}(\alpha)$ peaks  at $\alpha=35$. (Here, $P_{\rm
 Lap}(\alpha)$ and $P_{\rm Jef}(\alpha)$  represent $P(\alpha)$ for
 Laplace's and Jeffrey's rules, respectively. ~)
 Although $P_{\rm Lap}(\alpha)$  and $P_{\rm Jef}(\alpha)$ peak at
 different values of $\alpha$, similar images for ${\hat{\cal Z}}(\theta)$ 
 are obtained with slightly different errors, because ${\cal
 Z}^{(\alpha)}(\theta)$ is almost independent of $\alpha$ [see
 Eq. (\ref{eqn:finalZ})].  We, thus,
 find  in the non-flattening case that the MEM gives most probable
 images that are almost independent of the prior information and is consistent
 with the FTM. 
\subsubsection{Flattening case}
\label{subsub:flattening}
\begin{figure}[!h!t!b]
 \begin{center}
 \includegraphics[height=65mm]{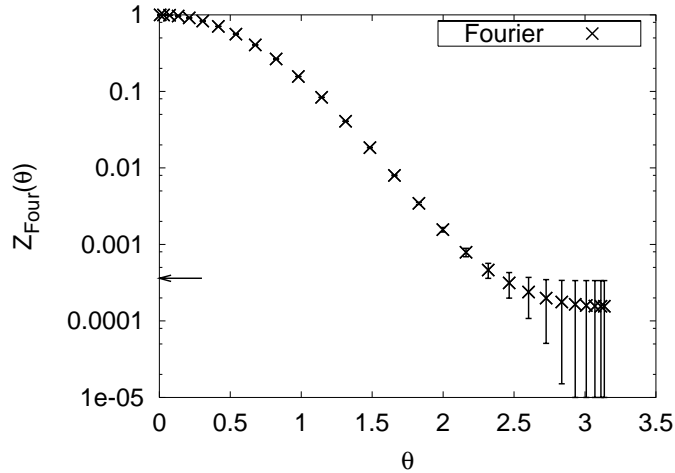}
 \end{center}\vspace*{0mm}
 \caption{Partition function  
 ${\cal Z}_{\rm Four}(\theta)$ obtained using the FTM. The number of
 measurements is 30.0M/set. The arrow indicates the value of  
 $\epsilon \hspace{.1cm}(=3.610\times 10^{-4})$. The errors were calculated with  the jackknife
 method.}
\label{fig:CP3FP-Fourier}
\end{figure}

 Now that we have found that the MEM is applicable to the analysis of the
 MC data, let us turn to the analysis of data with
 flattening ($L=50$). 
 In Fig. \ref{fig:CP3FP-Fourier}, we show ${\cal Z}_{\rm Four}(\theta)$
 obtained using  the FTM for data with 30.0M/set. Although ${\cal Z}_{\rm
 Four}(\theta)$ behaves 
 smoothly, its errors are large over a large range of values  of $\theta$ 
  (specifically for $\theta\simr
 2.4$). These large errors result from the error propagation of $P(Q)$
 through the Fourier transform. Let us here consider its effect by studying a
 quantity which represents the error propagation of $P(Q)$ in the case that
 there is no correlation of the distribution $P(Q)$ for  different values of
 $Q$, 
\begin{equation}
\epsilon\equiv\sum_Q |\Delta P(Q)|. \label{eqn:epsilon}
\end{equation}
 Here, $\epsilon=3.610\times 10^{-4}$,  and the value of ${\cal
 Z}_{\rm Four}(\theta)$ is comparable with that of $\epsilon$ at
 $\theta\simeq 2.4$. Figure \ref{fig:CP3FP-Fourier} displays that when
 the value of ${\cal 
 Z}_{\rm Four}(\theta)$  is smaller than that of $\epsilon$, the error
 in  ${\cal Z}_{\rm Four}(\theta)$ becomes large. This approximately holds
 for all cases we have investigated. In the MEM, $\epsilon$ could be an
 indicator for  the influence of the error of $P(Q)$ on  ${\hat{\cal
 Z}}(\theta)$.  This point  is  discussed in the following.
 In the analysis of data with
 flattening, much care is required.\cite{rf:ISY}
\par
 In order to properly evaluate  ${\hat{\cal Z}}(\theta)$ obtained using  the
 MEM,  we carefully 
 investigate (i) the statistical fluctuations of ${\hat{\cal
 Z}}(\theta)$,  (ii) the $g(\alpha)$ dependence of ${\hat{\cal Z}}(\theta)$, 
 and (iii) the relative error in  ${\hat{\cal Z}}(\theta)$. 
\\
 \uline{(i) The statistical fluctuations  of ${\hat{\cal Z}}(\theta)$}\par 
\begin{figure}[!h!t!b]
 \begin{center}
 \includegraphics[height=65mm]{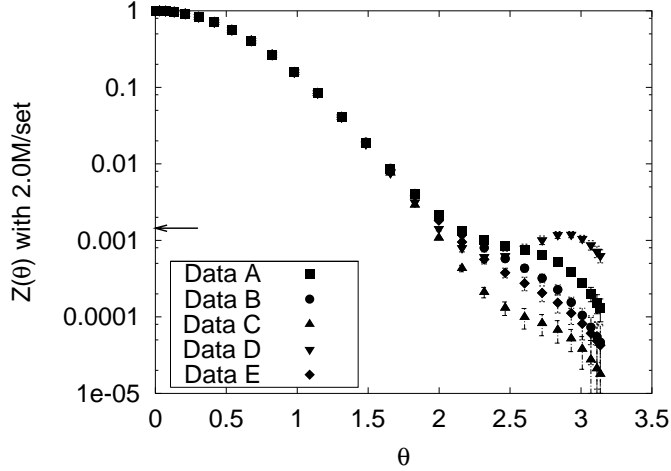}
 \end{center}\vspace*{0mm}
 \caption{The five most probable images ${\hat{\cal Z}}(\theta)$ for
 $L=50$ with 
 2.0M/set. Here, the Gaussian default model with $\gamma=5.0$ was  used.  The arrow 
 indicates the value of 
 $\epsilon\hspace{.1cm} (=1.441\times 10^{-3})$
 in Data A. The errors in ${\hat{\cal Z}}(\theta)$ were  calculated by  use of
 Eq. (\ref{eqn:uncertainty}).}
\label{fig:MEM_s2_g5_L50}
\end{figure}
 Figure \ref{fig:MEM_s2_g5_L50} displays five images 
 ${\hat{\cal Z}}(\theta)$, which are called Data A, B, C, D and E,
 respectively. These images were  calculated from the data with 2.0M/set, 
 which are independent of each other. The uncertainties in ${\hat{\cal
 Z}}(\theta)$ calculated with  Eq. (\ref{eqn:uncertainty}) are indicated as
 errors in Fig. \ref{fig:MEM_s2_g5_L50}. As a default model, the Gaussian
 default  with $\gamma=5.0$ were 
 used. It is seen that all the images ${\hat{\cal Z}}(\theta)$ fall on the same
 curve  for $\theta\siml 2.0$, while  they behave differently for $\theta
 \simr 2.0$.   The value of $\epsilon$ is at least $1.3\times
 10^{-3}$ for these five sets of  data,  and  the values of all the images 
 ${\hat{\cal Z}}(\theta)$  are smaller than that  of $\epsilon$ for
 $\theta\simr 2.0$. Figure \ref{fig:MEM_s2_g5_L50} indicates  
 that ${\hat{\cal Z}}(\theta)$ fluctuates greatly  when
 the value of ${\hat{\cal Z}}(\theta)$ is smaller than that of $\epsilon$. 
\par 
 To see  how ${\hat{\cal Z}}(\theta)$ depends on the statistics, we varied 
 the number of measurements.
 We fixed  the value of $\theta$ and chose the values
 $\theta=2.31~\mbox{and}~3.14$. The  value $\theta=2.31$ was   chosen for 
 the previously stated  reason that  ${\cal Z}_{\rm Four}(\theta)$ starts to
 contain  large 
 errors at this value for the data with 30.0M/set (see
 Fig. \ref{fig:CP3FP-Fourier}).  
\begin{figure}[!ht]\hspace*{-13mm}
 \begin{minipage}{.5\hsize}
  \begin{center}
   \includegraphics[height=60mm]{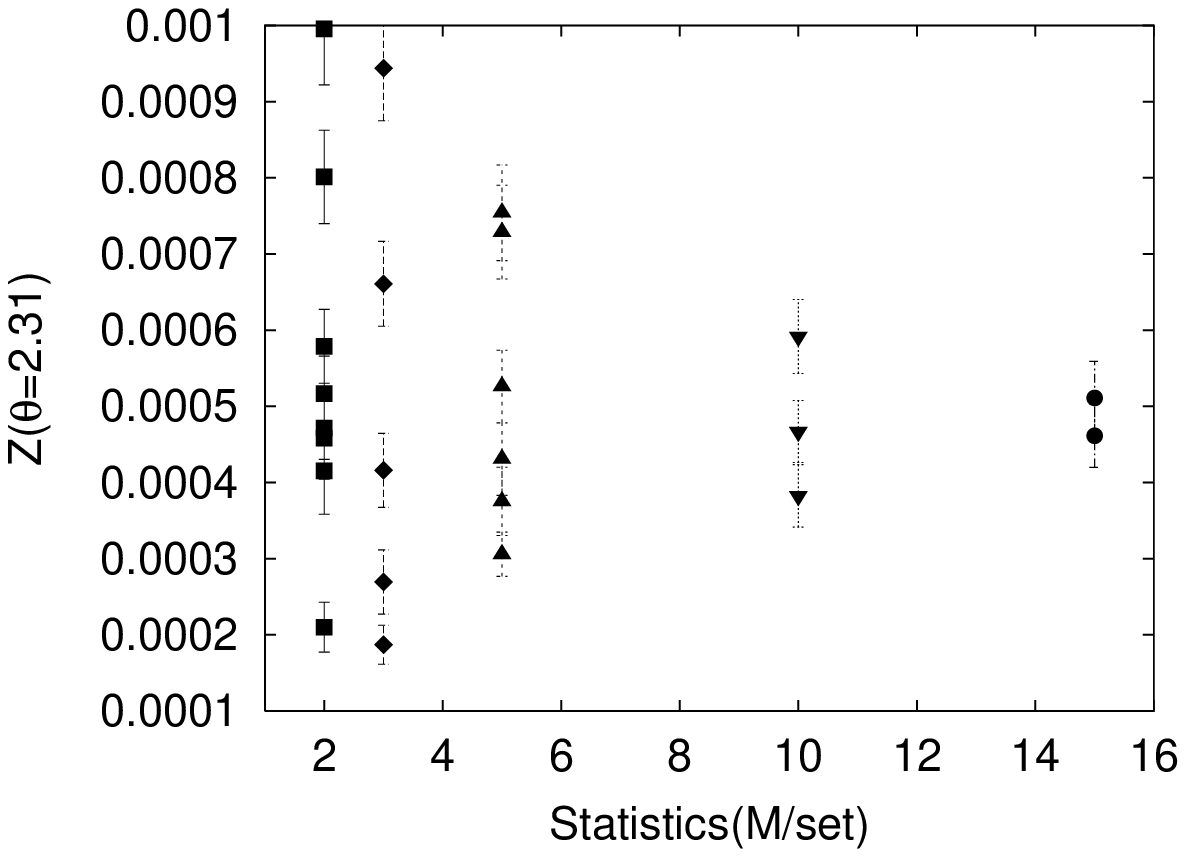}
  \end{center}
 \end{minipage}\hspace*{12mm}
 \begin{minipage}{.5\hsize}
  \begin{center}
   \includegraphics[height=60mm]{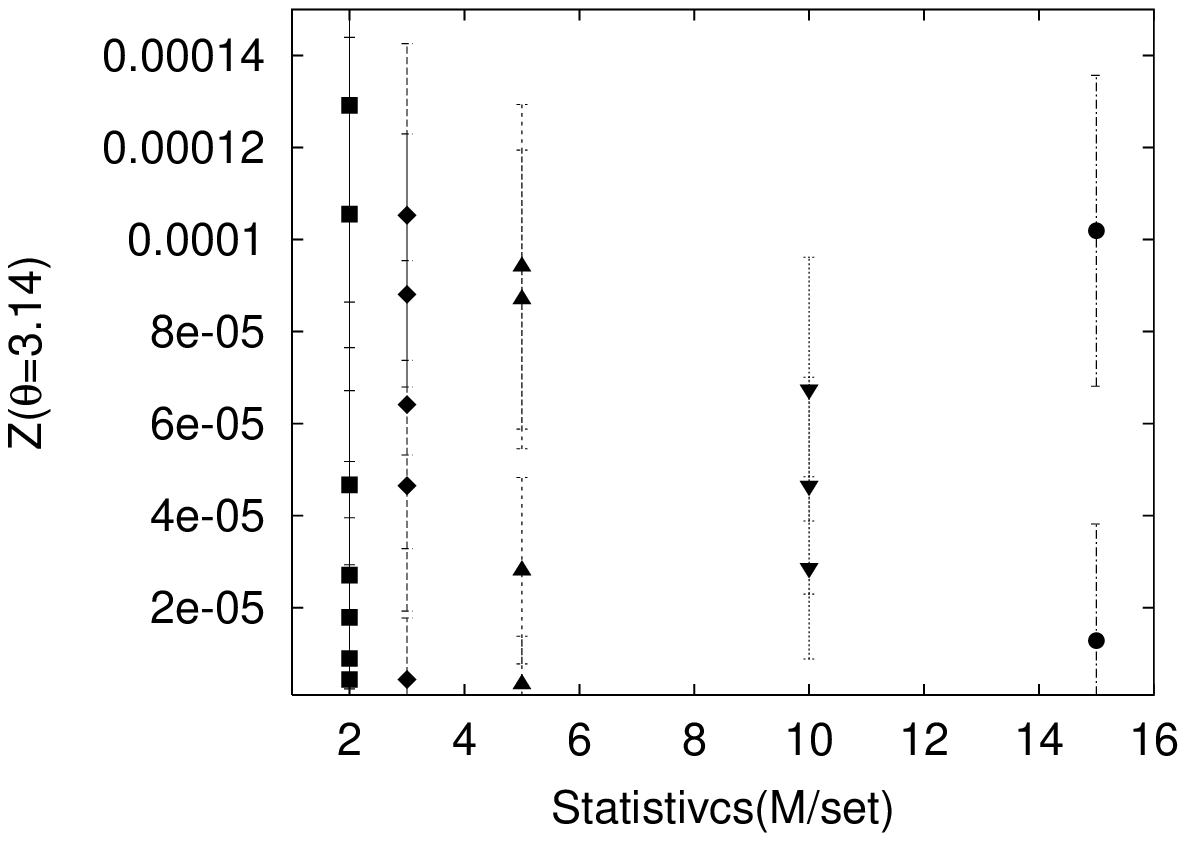}
  \end{center}
 \end{minipage} \vspace*{-1mm}
 \caption{Values of ${\hat{\cal Z}}(\theta)$ for $\theta=2.31$ (left
 panel) and 3.14 (right panel). The
 horizontal axis represents the number of measurements. Here, the Gaussian default
 model with $\gamma=5.0$ was  used.}
 \label{fig:MEM_g5_statistics}
\end{figure}
 Figure \ref{fig:MEM_g5_statistics} displays ${\hat{\cal Z}}(\theta)$ for 
 $\theta=2.31$ (left panel) and 3.14 (right panel). The horizontal axis
 represents the number of measurements. In the left panel, it is seen that the 
 fluctuations of ${\hat{\cal Z}}(2.31)$ become smaller as 
 the number of measurements increases. In the right panel, it is seen that 
  ${\hat{\cal Z}}(\theta)$ for
 15.0M/set fluctuates  as greatly as ${\hat{\cal Z}}(\theta)$ for
 2.0M/set. 
 In Table 
 \ref{table:MEM_g5_statistics}, the values of ${\hat{\cal Z}}(\theta)$
 for  $\theta=2.31,~2.60~\mbox{and}~3.14$ are listed for the cases of
 20.0M/set, 26.0M/set and 30.0M/set. For  $\theta=2.31$ and 2.60,  the images ${\hat{\cal
 Z}}(\theta)$ do not vary within the errors as the number of
 measurements increases. For $\theta=3.14$,  
 the value of ${\hat{\cal Z}}(\theta)$  varies significantly  as  the number 
 of measurements  increases. In the cases $\theta=2.31$ and 2.60, the values of
 ${\hat{\cal 
 Z}}(\theta)$ are comparable with that of $\epsilon$ for 30.0M/set
 ($\epsilon=3.610\times 10^{-4}$), while in the case
 $\theta=3.14$, the value of ${\hat{\cal 
 Z}}(\theta)$ is one order smaller than  $\epsilon$. This indicates that when
 the value of ${\hat{\cal Z}}(\theta)$ is smaller than that  of
 $\epsilon$, ${\hat{\cal Z}}(\theta)$ is strongly affected by 
 the error in $P(Q)$. Hereafter, we concentrate on the data with
 30.0M/set. 
%
\catcode`#=\active \def#{\phantom{0}}
\begin{table}[ht]
\caption{Values of ${\hat{\cal Z}}(\theta)$ for $\theta=2.31,~2.60$
 and $3.14$. Here, the same default model as in
 Fig. \ref{fig:MEM_g5_statistics} is used.}
\begin{center}
\begin{tabular}{c|ccc}
\hline
\hline
 statistics(M/set) & ${\hat{\cal Z}}_{\gamma=5.0}(2.31)$ 
                   & ${\hat{\cal Z}}_{\gamma=5.0}(2.60)$ 
                   & ${\hat{\cal Z}}_{\gamma=5.0}(3.14)$ \\
\hline
 20.0 & 4.07(39)$\times 10^{-4}$ & 2.30(31)$\times 10^{-4}$
      & 3.93(22)$\times 10^{-5}$ \\
 26.0 & 4.22(40)$\times 10^{-4}$ & 2.69(33)$\times 10^{-4}$
      & 5.83(26)$\times 10^{-5}$ \\
 30.0 & 4.58(40)$\times 10^{-4}$ & 2.81(33)$\times 10^{-4}$ 
      & 5.37(26)$\times 10^{-5}$\\
\hline
\end{tabular}
\end{center}
\label{table:MEM_g5_statistics}
\end{table}
\vspace*{2mm}\\
\uline{(ii) The $g(\alpha)$ dependence of ${\hat{\cal Z}}(\theta)$}
\par 
 The most probable image ${\hat{\cal Z}}(\theta)$ is obtained by
 performing the integral with respect to $\alpha$ according to  the procedure
 2  outlined in \S \ref{sub:MEM}. The probability $P(\alpha)$ in
 Eq. (\ref{eqn:posterioralpha}) involves  the prior
 probability of $\alpha$, $g(\alpha)$. In the present study, we 
 investigate the $g(\alpha)$ dependence of ${\hat{\cal Z}}(\theta)$ by
 calculating Eq. (\ref{eqn:Delta}).
\par
 Before calculating $\Delta(\theta)$, we investigate how $g(\alpha)$ affects the
 behavior of $P(\alpha)$. From the definition given  in
 \S\ref{sub:comment}, the following relation holds,  up to the
 normalization constant between $P_{\rm Lap}(\alpha)$ and $P_{\rm
 Jef}(\alpha)$:
\begin{equation}
 P_{\rm Jef}(\alpha)\propto g_{\rm Jef}(\alpha)P_{\rm Lap}(\alpha)
  =\frac{1}{\alpha}P_{\rm Lap}(\alpha). \label{eqn:Pjef-Plap}
\end{equation}
 The probability $g_{\rm Jef}(\alpha)$ deforms the shape of $P_{\rm
 Lap}(\alpha)$ and shifts the location of its
 peak.\cite{rf:Yamazaki_CP-PACS} It was shown in 
 Ref. \citen{rf:AHN} that when $P_{\rm
 Lap}(\alpha)$ is concentrated around its maximum at
 $\alpha=\hat\alpha$,  the peaks of $P_{\rm Lap}(\alpha)$ 
 and $P_{\rm Jef}(\alpha)$ are located at nearly equal values  of $\alpha$,  and
 ${\hat{\cal Z}}(\theta)$ is insensitive to the choice of $g(\alpha)$.
 We  quantitatively determine the amount by which  $g_{\rm Jef}(\alpha)$ shifts the location
 of the  peak of $P_{\rm Lap}(\alpha)$ in the following.
 \par
 The derivative of $P(\alpha)$ with respect to $\alpha$,  
\begin{eqnarray}
 \frac{d P(\alpha)}{d\alpha}&=&\left[
  \frac{1}{2\alpha}\sum_k\frac{\lambda_k(\alpha)}{\alpha+\lambda_k(\alpha)}
  +S(\alpha)+\frac{d\ln g(\alpha)}{d\alpha}\right. \nonumber \\ 
 &&\left.-\frac{1}{2}\frac{d\chi^2(\alpha)}{d\alpha}+
 \alpha\frac{dS(\alpha)}{d\alpha} 
   -\frac{1}{2}\sum_k
   \frac{1}{\alpha+\lambda_k(\alpha)}\frac{d\lambda_k(\alpha)}{d\alpha}
			       \right]P(\alpha), 
   \label{eqn:extremepoint}
\end{eqnarray}
 is vanishing at $\alpha={\hat\alpha}$. It is noted that
 $\chi^2(\alpha)$, $S(\alpha)$ and $\{\lambda_k(\alpha)\}$ depend
 implicitly on $\alpha$ through ${\cal Z}^{(\alpha)}(\theta)$,  calculated
 using  Eq. (\ref{eqn:Zalp}). We obtain
\begin{eqnarray}
 {\hat\alpha}_{\rm Lap}S({\hat\alpha}_{\rm Lap})&=& -\frac{1}{2}N_g +
  \mbox{[derivative terms],}\hspace*{11mm}\mbox{(Laplace's rule)}
  \label{eqn:Gullcriterion1} \\
 {\hat\alpha}_{\rm Jef}S({\hat\alpha}_{\rm Jef})&=& -\frac{1}{2}(N_g-2) +
  \mbox{[derivative terms],}\hspace*{3mm}\mbox{(Jeffrey's rule)}
  \label{eqn:Gullcriterion2} 
\end{eqnarray}
 where ``derivative terms'' represents the derivatives of
 $\chi^2(\alpha),~S(\alpha)$ and $\lambda_k(\alpha)$, and
 ${\hat\alpha}_{\rm Lap}$ and ${\hat\alpha}_{\rm Jef}$ denote 
 ${\hat\alpha}$ for Laplace's and Jeffrey's rules, respectively. 
 Here, 
 $N_g\equiv\sum_k\lambda_k({\hat\alpha})/({\hat\alpha}+\lambda_k({\hat\alpha}))$,
 and $N_g$ represents the number of effectively independent measurements, 
 because the value of the $\lambda_k({\hat\alpha})$ contributes
 approximately 1  to the summation when
 $\lambda_k({\hat\alpha})\gg{\hat\alpha}$. Note that since
 the quantities $\lambda_k(\alpha)$ are independent of the choice of $g(\alpha)$,
 $N_g$ is also insensitive to $g(\alpha)$. 
 For simplicity, let us ignore these ``derivative terms''  for the time being.
   With the relative difference between the two values of  ${\hat\alpha}$, 
\begin{equation}
 r_{{\hat\alpha}}\equiv
  \frac{{\hat\alpha}_{\rm Lap}-{\hat\alpha}_{\rm Jef}}{{\hat\alpha}_{\rm
  Lap}}, \label{eqn:diffalpha_0}
\end{equation}
 we obtain, from Eqs. (\ref{eqn:Gullcriterion1}) and (\ref{eqn:Gullcriterion2}), 
\begin{equation}
 r_{{\hat\alpha}} = \frac{2}{N_g}, \label{eqn:diffalpha}
\end{equation}
 because the derivative of $S(\alpha)$ is vanishing and
 $S({\hat\alpha}_{\rm Lap})\simeq S({\hat\alpha}_{\rm Jef})$. 
 Hence, the relative
 difference between the values of  ${\hat\alpha}$, $r_{{\hat\alpha}}$, to good approximation
  depends  only on 
 $N_g$. When $N_g$ is 
 larger than 2, $r_{{\hat\alpha}}$ is
 negligible,  and the peaks of $P_{\rm Lap}(\alpha)$ and $P_{\rm Jef}(\alpha)$ could
 be located at almost equal values   of $\alpha$. In case that the
 ``derivative terms'' cannot be ignored, however, using only $N_g$,  one cannot properly estimate
  to what extent  the location of the 
 peak moves. In such a case, we need to resort to numerical
 calculations. 
\par
 Let us investigate the behavior of   $P(\alpha)$  in our data;  specifically, we check
 whether the derivatives of $\chi^2(\alpha)$, $S(\alpha)$ and
 $\{\lambda_k(\alpha)\}$ are negligible and determine the distance between   the peaks of 
 $P_{\rm Lap}(\alpha)$ and $P_{\rm Jef}(\alpha)$. The values of
 ${\hat\alpha}$, ${\hat\alpha}S({\hat\alpha})$, $N_g/2$ and 
 $|{\hat\alpha}S({\hat\alpha})+N_g/2|\equiv D$ for various $m_{\rm
 G}(\theta)$ are listed in Table \ref{table:extremepoint}  in the case of Laplace's
 rule. As shown in Eq. (\ref{eqn:Gullcriterion1}), 
 the value of $D$ provides  a criterion for determining  whether or not  the ``derivative terms'' 
 can be ignored. 
\par
\begin{figure}[!th]\hspace*{-13mm}
 \begin{minipage}{.5\hsize}
  \begin{center}
   \includegraphics[height=60mm]{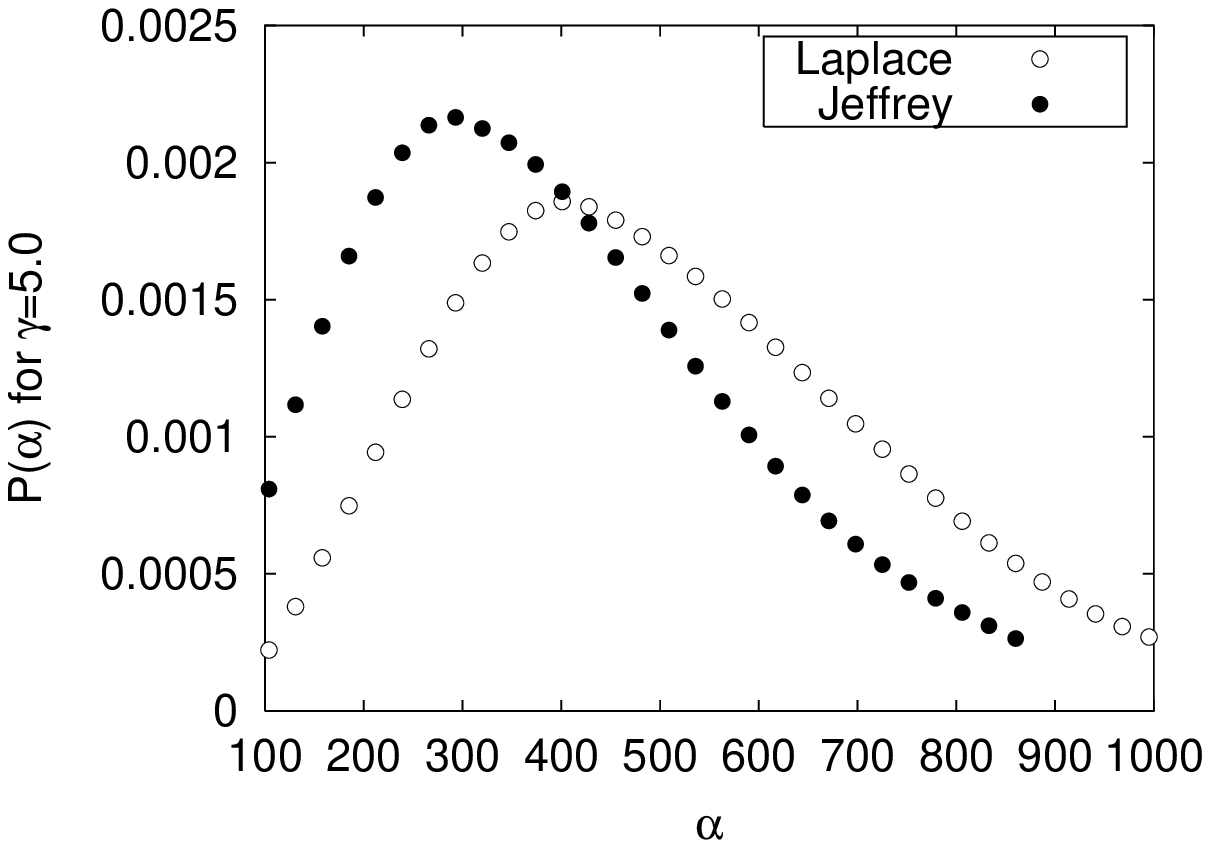}
  \end{center}
 \end{minipage}\hspace*{12mm}
 \begin{minipage}{.5\hsize}
  \begin{center}
   \includegraphics[height=60mm]{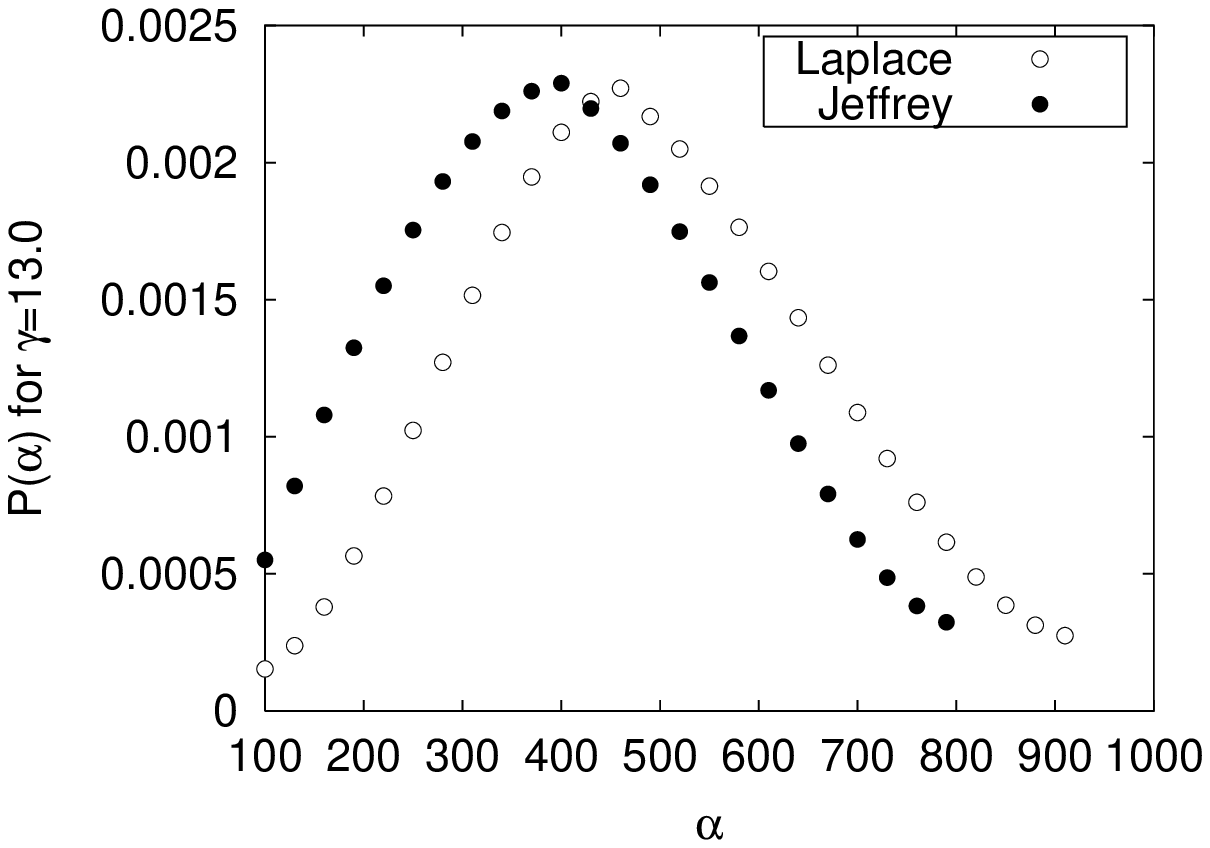}
  \end{center}
 \end{minipage} \vspace*{-1mm}
 \caption{$P_{\rm Lap}(\alpha)$ and
 $P_{\rm Jef}(\alpha)$. As the default models,  Gaussian functions with
 $\gamma=5.0$ (left panel) and $13.0$ (right panel) were  used.}
 \label{fig:Palpha_LJ}
\end{figure}

 Firstly, we focus on the $\gamma=5.0$ case. In
 Eq. (\ref{eqn:Gullcriterion1}), $D$ makes a smaller contribution to
 ${\hat\alpha}_{\rm Lap}S({\hat\alpha}_{\rm Lap})$ than does   $N_g/2$ 
 ($D\simeq 0.04\times N_g/2$),  
 and hence the ``derivative terms'' can be ignored. It is thus  expected
 that Eq. (\ref{eqn:diffalpha}) holds. Substituting ${\hat\alpha}_{\rm
 Lap}$ and $N_g$ in Table \ref{table:extremepoint} into
 Eq. (\ref{eqn:diffalpha}),  we obtain ${\hat\alpha}_{\rm Jef}\simeq
 287$. With  Jeffrey's rule, on the other hand, $P_{\rm Jef}(\alpha)$
 gives ${\hat\alpha}_{\rm Jef}=293$. We thus find   good agreement between the values of  
 ${\hat\alpha}_{\rm Jef}$ obtained in these two ways. Next, we determine the distance  between  the peaks of
 $P_{\rm Lap}(\alpha)$ and $P_{\rm Jef}(\alpha)$. Because  the
 value of $N_g$ is comparable to 2 $(r_{{\hat\alpha}}\simeq 0.27)$, it
 is expected that the peaks of $P_{\rm Lap}(\alpha)$ and $P_{\rm Jef}(\alpha)$ could
  appear at somewhat separated  values of $\alpha$. 
   It is shown that this is indeed the case  in the
 left panel of Fig. \ref{fig:Palpha_LJ}.
\par
 Now let us consider  the other cases  of the default model  $m_{\rm G}(\theta)$
   in Table \ref{table:extremepoint}. The values of $D$ are larger
 than that  for $\gamma=5.0$: we have  $D\simeq
 0.67\times N_g/2$, $D\simeq
 0.18\times N_g/2$ and $D\simeq
 0.30\times N_g/2$ for
 $\gamma=8.0$, 10.0 and 13.0, 
 respectively. In these cases, the ``derivative terms''  are not
 negligible,  and Eq. (\ref{eqn:diffalpha}) no longer  holds. The
 probabilities $P(\alpha)$ for Laplace's and Jeffrey's rules give 
 $r_{{\hat\alpha}}\simeq 0.20,$ 0.14 and 0.13 for
 $\gamma=8.0$, 10.0 and 13.0, respectively. The right panel of
 Fig. \ref{fig:Palpha_LJ} displays both  $P_{\rm Lap}(\alpha)$ and
 $P_{\rm Jef}(\alpha)$ for $\gamma=13.0$. In this figure,
  the peaks of $P_{\rm Lap}(\alpha)$ and $P_{\rm Jef}(\alpha)$
  are seen to be located   at different values of $\alpha$. Similar behavior 
 of $P(\alpha)$ is  obtained for the other default models $m(\theta)$. 
\par
\begin{table}[ht!]
\caption{Values of
 ${\hat\alpha}_{\rm Lap},~{\hat\alpha}_{\rm Lap}S({\hat\alpha}_{\rm
 Lap}),-N_g/2$ and  
 $D(\equiv |{\hat\alpha}_{\rm Lap}S({\hat\alpha}_{\rm Lap})+N_g/2|)$ 
 for various
 $m_{\rm G}(\theta)$ with  Laplace's rule. Here, $N_Q=7$. }
\begin{center}
\begin{tabular}{c|ccccc}
\hline
\hline
 $\gamma$(Gaussian default) & ${\hat\alpha}_{\rm Lap}$ &
 ${\hat\alpha}_{\rm Lap}S({\hat\alpha}_{\rm Lap})$ & $-N_g/2$ & 
 $|{\hat\alpha}_{\rm Lap}S({\hat\alpha}_{\rm Lap})+N_g/2|\equiv D$ \\ 
\hline
 #5.0 & 401 & $-3.358$ & $-3.498$ & 0.140($\simeq
 0.04\times N_g/2$) \\ 
 #8.0 & 10200 & $-0.992$ & $-3.046$ & 2.054($\simeq
 0.67\times N_g/2$) \\ 
 10.0 & 1400 & $-2.653$ & $-3.232$ & 0.579($\simeq
 0.18\times N_g/2$) \\  
 13.0 & 460 & $-4.419$ & $-3.410$ & 1.009($\simeq
 0.30\times N_g/2$) \\ 
\hline
\end{tabular}
\end{center}
\label{table:extremepoint}
\end{table}
\par
\begin{figure}[!th]\hspace*{-13mm}
 \begin{minipage}{.5\hsize}
  \begin{center}
   \includegraphics[height=60mm]{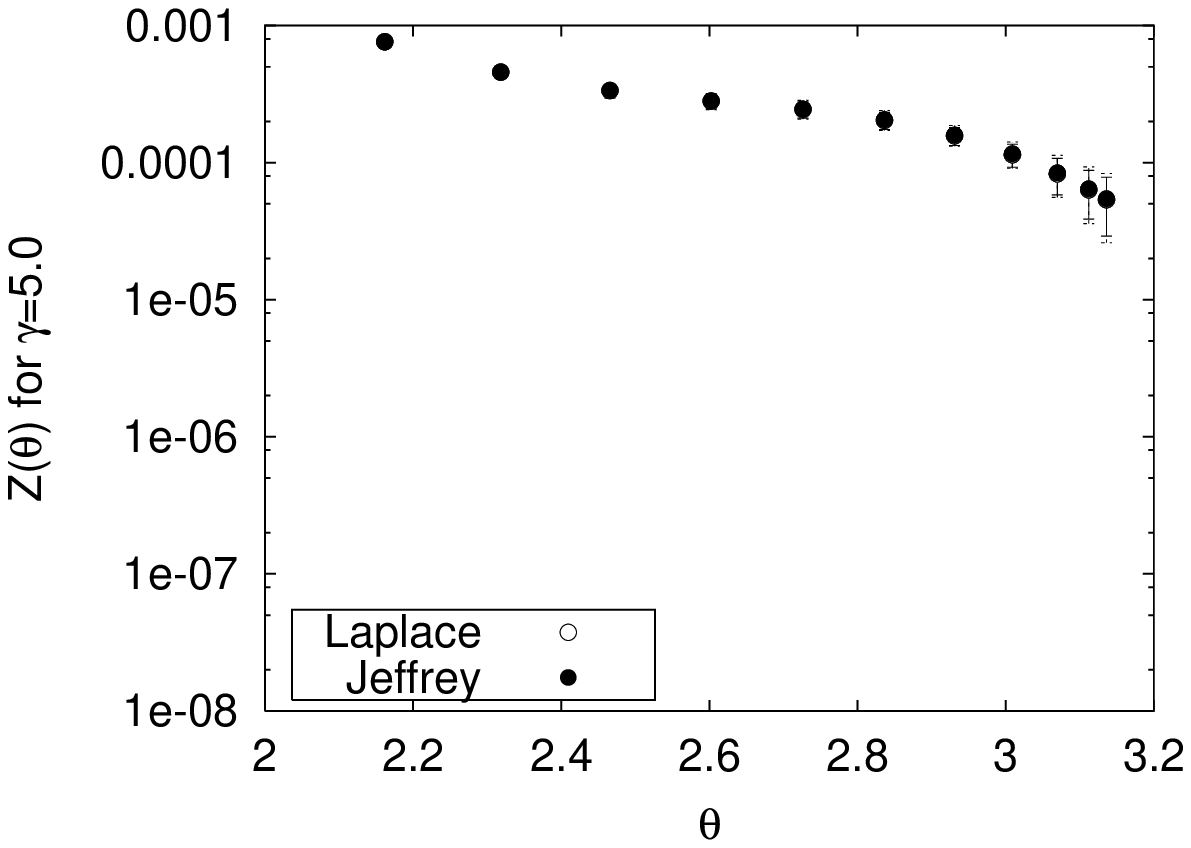}
  \end{center}
 \end{minipage}\hspace*{12mm}
 \begin{minipage}{.5\hsize}
  \begin{center}
   \includegraphics[height=60mm]{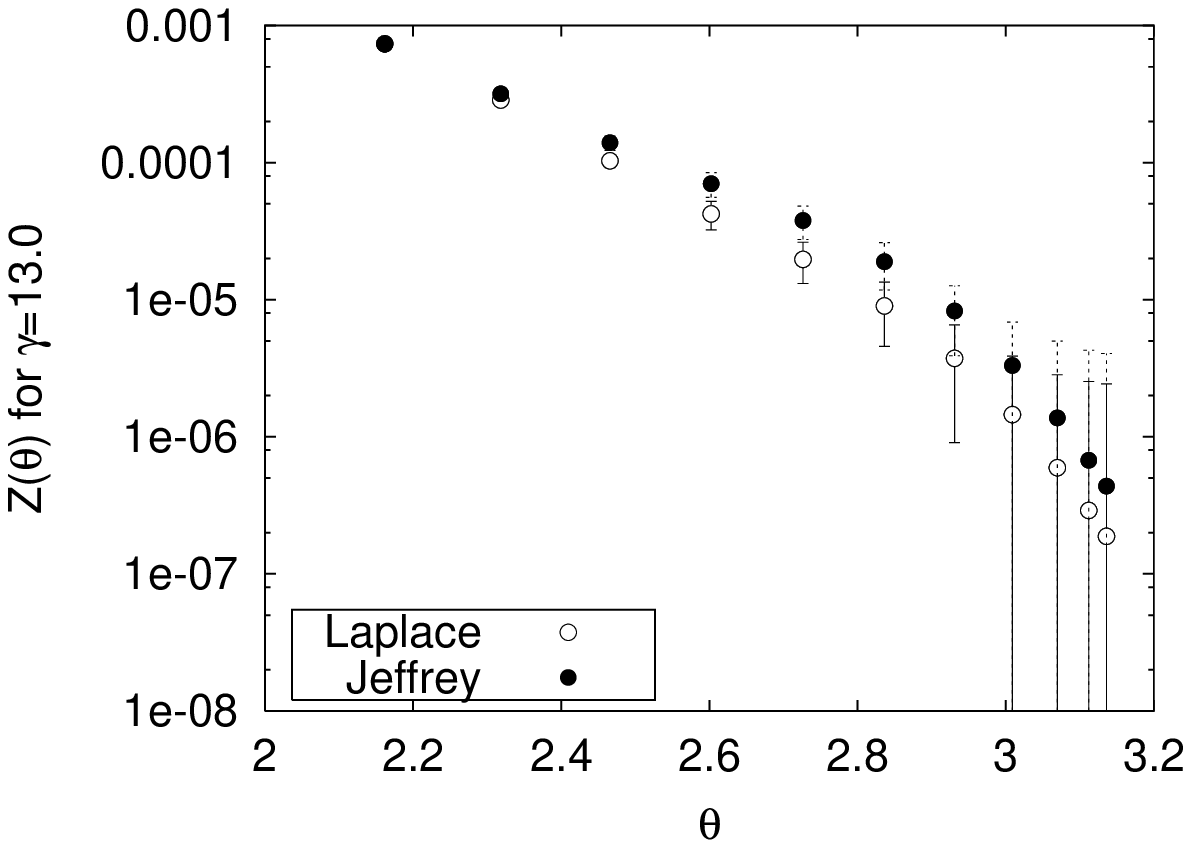}
  \end{center}
 \end{minipage} \vspace*{-1mm}
 \caption{${\hat{\cal Z}}_{\rm Lap}(\theta)$ and ${\hat{\cal
 Z}}_{\rm Jef}(\theta)$ for $\theta\in[2.0,\pi]$. The  default models used here are the same 
 as in Fig. \ref{fig:Palpha_LJ}. The errors in  ${\hat{\cal
 Z}}(\theta)$ were calculated using  Eq. (\ref{eqn:uncertainty}).}
 \label{fig:finalZ_LJ}
\end{figure}

 Let us now study   ${\hat{\cal Z}}(\theta)$. The
 $\alpha$-dependent image ${\cal Z}^{(\alpha)}(\theta)$ affects the
 behavior of ${\hat{\cal Z}}(\theta)$ through the integral in
 Eq. (\ref{eqn:finalZ}):
%
\begin{eqnarray}
 {\hat{\cal Z}}(\theta)=\left\{
   \begin{array}{l}
 \int d\alpha~P_{\rm Lap}(\alpha){\cal Z}^{(\alpha)}(\theta), 
  \mbox{\hspace*{4mm}(Laplace's rule)} \\
 \int d\alpha~P_{\rm Jef}(\alpha){\cal Z}^{(\alpha)}(\theta).
  \mbox{\hspace*{6.5mm}(Jeffrey's rule)}
  \label{eqn:finalZ_LJ}
   \end{array}\right.
\end{eqnarray}
 If ${\cal Z}^{(\alpha)}(\theta)$ does not vary over the
 range of integration in Eq. (\ref{eqn:finalZ_LJ}), ${\cal
 Z}^{(\alpha)}(\theta)$ can be factored out from the integral and we have 
 ${\hat{\cal Z}}_{\rm Lap}(\theta)={\hat{\cal Z}}_{\rm
 Jef}(\theta)$,  due to the normalization of $P(\alpha)$.   
 This is indeed  the case for $\gamma=5.0$: for  $\theta=2.60$, for example,
 the values of ${\cal Z}^{(\alpha)}(\theta)$ are $2.831\times 10^{-4}$
 for $\alpha=50$ and $2.797\times 10^{-4}$ for $\alpha=1050$. The latter 
 value of $\alpha$ is the upper limit of $\alpha$, $\alpha_{\rm max}$,
 and the former one is the lower limit of $\alpha$,  $\alpha_{\rm min}$,
 in the integral. For the other default  models 
 listed in Table \ref{table:extremepoint}, by contrast, all the ${\cal
 Z}^{(\alpha)}(\theta)$  vary over several orders for $\theta\simr
 2.6$.  For $\gamma=13.0$, for example,  the values of
 ${\cal Z}^{(\alpha)}(2.60)$ are $2.80\times 10^{-4}$
 and $9.55\times 10^{-7}$ for $\alpha=\alpha_{\rm min}=100$ and
 $\alpha=\alpha_{\rm max}=910$, respectively.    
 From these results, it is conjectured  that ${\hat{\cal Z}}(\theta)$ is
 almost independent of $g(\alpha)$ for $\gamma=5.0$ and depends strongly 
 on $g(\alpha)$  for the other values of $\gamma$. In fact, no
 $g(\alpha)$ dependence is seen over the entire the range of values   of $\theta$ for
 $\gamma=5.0$,  as shown in the left panel of  Fig. \ref{fig:finalZ_LJ},
 while in the right panel (depicting the situation for $\gamma=13.0$),    
 differences between these two are observed for $\theta\simr 2.6$. 
 Note that for $\theta\siml
 2.0$, ${\hat{\cal Z}}_{\rm Lap}(\theta)$ and ${\hat{\cal Z}}_{\rm
 Jef}(\theta)$ fall on the same curve in both the $\gamma=5.0$ and 13.0
 cases.   \par 
\par
\begin{figure}
 \begin{center}
 \includegraphics[height=70mm]{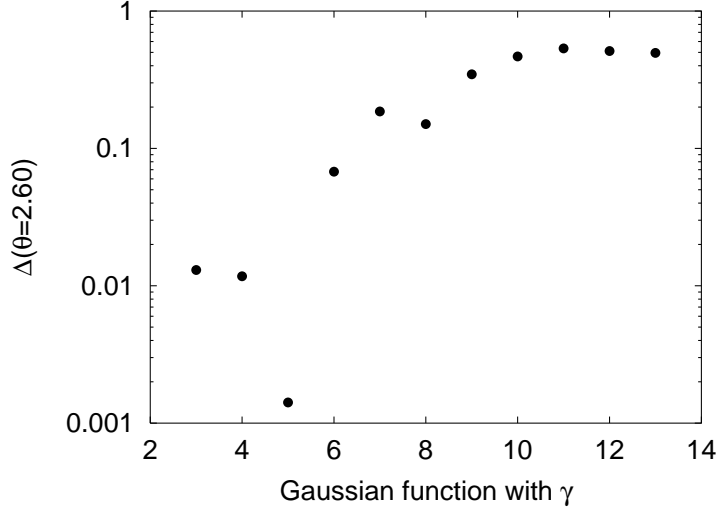}
\end{center}
\caption{Values of $\Delta(\theta)$ for $\theta=2.60$. The Gaussian
 functions are used as the  default models. The horizontal axis represents 
 the value of $\gamma$ in $m_{\rm G}(\theta)$.}
\label{fig:Delta}
\end{figure}

 In order to estimate the influence of $g(\alpha)$ on ${\hat{\cal
 Z}}(\theta)$, we calculate $\Delta(\theta)$ given by 
 Eq. (\ref{eqn:Delta}). 
 For a systematic estimation, the parameter $\gamma$ in 
 $m_{\rm G}(\theta)$ is varied from 3.0 to 13.0. Figure
 \ref{fig:Delta} displays the values of $\Delta(\theta)$ at
 $\theta=2.60$. The horizontal axis represents  the value of
 $\gamma$. It is found that the value of $\Delta(\theta)$ is  smallest for
 $\gamma=5.0$ and becomes larger as the value of
 $\gamma$ deviates from 5.0.
 Table \ref{table:Deltavalues} lists the values of
 $\Delta(\theta)$ for eight images among these
  with  $\theta=2.31,~2.60,~2.83~\mbox{and}~3.14$. 
 In the case of  $\gamma=5.0$,  the 
 value of $\Delta(\theta)$ increases  with  $\theta$ for $\theta\simr 2.6$. 
  This behavior is also observed  for   the  other
 $m(\theta)$.  It is 
 found that the values of $\Delta(\theta)$ for $\gamma=3.0$, 4.0, 5.0, 
 $m_{24/50}(\theta)$, $m_{32/50}(\theta)$ and $m_{38/50}(\theta)$ are
  quite  small over the entire range of  values  of $\theta$ and that the images 
 ${\cal Z}^{(\alpha)}(\theta)$  for these six depend  only very weakly on
 $\alpha$ over the range of integration in Eq. (\ref{eqn:finalZ}). 
\par
 We now give a brief  comment on $\epsilon$ in Eq. (\ref{eqn:epsilon}). 
 The difference between ${\hat{\cal Z}}_{\rm Lap}(\theta)$ and
 ${\hat{\cal Z}}_{\rm Jef}(\theta)$ becomes
 significant at the value  $\theta=\theta_{\epsilon}$, where
 ${\hat{\cal Z}}(\theta_{\epsilon})\simeq\epsilon$ is satisfied, and
 $\Delta(\theta_{\epsilon})\simeq 0.1$ holds; e.g., for
 $\gamma=13.0$, we have $\theta_{\epsilon}\simeq 2.3$ (see
 Fig. \ref{fig:finalZ_LJ}).  This 
 is true for all  cases in which  $\gamma\ge 7.0$. 
\par 
\begin{table}[ht]
\caption{Values of $\Delta (\theta)$ at
 $\theta=2.31,2.60,2.83$ and 3.14 for various $m(\theta)$.}
\begin{center}
\begin{tabular}{c||cccc}
\hline
\hline
default model & $\Delta (2.31)$ & $\Delta (2.60)$ & $\Delta (2.83)$ &
 $\Delta (3.14)$ \\ 
\hline
$m_{\rm G}(\theta)$ with $\gamma=3.0$  & 6.30$\times$10$^{-3}$ &
 1.30$\times$10$^{-2}$ & 1.87$\times$10$^{-2}$ & 2.24$\times$10$^{-2}$ \\
$m_{\rm G}(\theta)$ with $\gamma=5.0$ & 5.35$\times$10$^{-3}$ &
 1.41$\times$10$^{-3}$ & 1.12$\times$10$^{-2}$ & 1.84$\times$10$^{-2}$ \\
$m_{\rm G}(\theta)$ with $\gamma=8.0$ & 1.03$\times$10$^{-1}$ &
 1.50$\times$10$^{-1}$ & 1.70$\times$10$^{-1}$ & 1.75$\times$10$^{-1}$ \\
$m_{\rm G}(\theta)$ with $\gamma=10.0$ & 1.01$\times$10$^{-1}$ &
 4.67$\times$10$^{-1}$ & 8.21$\times$10$^{-1}$ & 9.60$\times$10$^{-1}$ \\
$m_{\rm G}(\theta)$ with $\gamma=13.0$ & 1.05$\times$10$^{-1}$ & 
 4.96$\times$10$^{-1}$ & 7.12$\times$10$^{-1}$ & 7.95$\times$10$^{-1}$ \\
$m_{24/50}(\theta)$ & 3.44$\times$10$^{-3}$ &
 7.67$\times$10$^{-3}$ & 1.13$\times$10$^{-2}$ & 1.36$\times$10$^{-2}$ \\
$m_{32/50}(\theta)$ & 7.30$\times$10$^{-3}$ &
 1.57$\times$10$^{-2}$ & 2.29$\times$10$^{-2}$ & 2.72$\times$10$^{-2}$ \\
$m_{38/50}(\theta)$ & 1.30$\times$10$^{-2}$ &
 2.60$\times$10$^{-2}$ & 3.65$\times$10$^{-2}$ & 4.26$\times$10$^{-2}$ \\
\hline
\end{tabular}
\end{center}
\label{table:Deltavalues}
\end{table}
\par
\begin{figure}[!ht]
 \begin{center}
 \includegraphics[height=70mm]{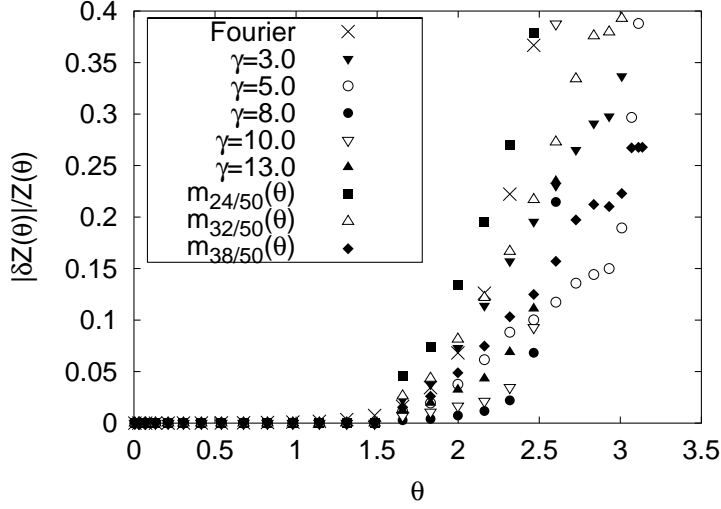}
\end{center}
\caption{Values of $|\delta {\hat{\cal Z}}(\theta)|/
{\hat{\cal Z}}(\theta)$ for the selected default models. The results obtained with the Fourier method 
  are  also  plotted.
}
\label{fig:relativeerror}
\end{figure}
\begin{table}[ht]
\caption{Values of $|\delta{\hat{\cal Z}}|/{\hat{\cal Z}}(\theta)\equiv 
|\delta{\hat{\cal Z}}(\theta)|/{\hat{\cal Z}}(\theta)$ for 
 $\theta=2.31,2.60,2.83$ and 3.14. Here, the same default models as in Table
 \ref{table:Deltavalues}  were  used.}
\begin{center}
\begin{tabular}{c||cccc}
\hline
\hline
default model & $|\delta{\hat{\cal Z}}|/{\hat{\cal Z}}(2.31)$ &
 $|\delta{\hat{\cal Z}}|/{\hat{\cal Z}}(2.60)$ & 
 $|\delta{\hat{\cal Z}}|/{\hat{\cal Z}}(2.83)$ & 
 $|\delta{\hat{\cal Z}}|/{\hat{\cal Z}}(3.14)$ \\ 
\hline
$m_{\rm G}(\theta)$ with $\gamma=3.0$  & 1.57$\times$10$^{-1}$ &
 2.31$\times$10$^{-1}$ & 2.91$\times$10$^{-1}$ & 6.24$\times$10$^{-1}$ \\
$m_{\rm G}(\theta)$ with $\gamma=5.0$ & 8.82$\times$10$^{-2}$ &
 1.17$\times$10$^{-1}$ & 1.44$\times$10$^{-1}$ & 4.58$\times$10$^{-1}$ \\
$m_{\rm G}(\theta)$ with $\gamma=8.0$ & 2.21$\times$10$^{-2}$ &
 2.15$\times$10$^{-1}$ & 1.93 & 56.84 \\
$m_{\rm G}(\theta)$ with $\gamma=10.0$ & 3.48$\times$10$^{-2}$ &
 3.88$\times$10$^{-1}$ & 1.86 & 40.69 \\
$m_{\rm G}(\theta)$ with $\gamma=13.0$ & 6.88$\times$10$^{-2}$ & 
 2.35$\times$10$^{-1}$ & 4.93$\times$10$^{-1}$ & 11.94 \\
$m_{24/50}(\theta)$ & 2.70$\times$10$^{-1}$ &
 4.82$\times$10$^{-1}$ & 6.54$\times$10$^{-1}$ & 8.31$\times$10$^{-1}$ \\
$m_{32/50}(\theta)$ & 1.67$\times$10$^{-1}$ &
 2.73$\times$10$^{-1}$ & 3.76$\times$10$^{-1}$ & 4.82$\times$10$^{-1}$ \\
$m_{38/50}(\theta)$ & 1.03$\times$10$^{-1}$ &
 1.57$\times$10$^{-1}$ & 2.12$\times$10$^{-1}$ & 2.68$\times$10$^{-1}$ \\
\hline
\end{tabular}
\end{center}
\label{table:relativevalues}
\end{table}
%
\uline{(iii) The relative error of ${\hat{\cal Z}}(\theta)$}\par
 Now that  the 
 $g(\alpha)$ dependence of ${\hat{\cal Z}}(\theta)$ has been
 systematically investigated,    
 we next consider the uncertainty  in ${\hat{\cal Z}}(\theta)$. 
\par
 The relative errors in  ${\hat{\cal Z}}(\theta)$, $|\delta{\hat{\cal
 Z}}(\theta)|/{\hat{\cal Z}}(\theta)$, are displayed 
 in Fig. \ref{fig:relativeerror}, where $\delta{\hat{\cal Z}}(\theta)$
 is calculated using  Eq. (\ref{eqn:uncertainty}). For  comparison,  
 $|\delta{\cal Z}_{\rm Four}(\theta)|/{\cal Z}_{\rm Four}(\theta)$ 
 is also plotted. It is observed that all the relative errors increase with
  $\theta$.  In particular, those for $\gamma=8.0$, 10.0 and 13.0
 diverge  for  large $\theta$  ($\theta\simr 2.6$). The value of 
 $|\delta{\hat{\cal Z}}(\theta)|/{\hat{\cal Z}}(\theta)$ for 
 $m_{24/50}(\theta)$ is comparable with that of $|\delta{\cal Z}_{\rm
 Four}(\theta)|/{\cal Z}_{\rm Four}(\theta)$,  and those for the others
 are smaller than that  of $|\delta{\cal Z}_{\rm Four}(\theta)|/{\cal
 Z}_{\rm Four}(\theta)$. 
 To see in detail how
 $|\delta{\hat{\cal Z}}(\theta)|/{\hat{\cal Z}}(\theta)$ varies in the 
 large $\theta$ region, we list $|\delta{\hat{\cal
 Z}}(\theta)|/{\hat{\cal Z}}(\theta)$ with  $\theta=2.31$, 2.60, 2.83 and
 3.14 in Table \ref{table:relativevalues} for various $m(\theta)$. The
 relative errors $|\delta{\hat{\cal Z}}(\theta)|/{\hat{\cal Z}}(\theta)$
 for $\gamma=8.0$, 10.0 and 13.0 increase  rapidly  with  
 $\theta$ and exceed 1.0 for $\theta\simr 2.8$.
 It is seen in Table \ref{table:relativevalues} that 
  for $\gamma=5.0$,
 $m_{32/50}(\theta)$ and $m_{38/50}(\theta)$,  the images ${\hat{\cal Z}}(\theta)$ have 
   small uncertainties.   
\par
 As seen in (ii), some of the images ${\cal Z}^{(\alpha)}(\theta)$ vary over several orders
  as  $\alpha$ varies, and the images ${\hat{\cal Z}}(\theta)$ calculated from
 these ${\cal Z}^{(\alpha)}(\theta)$ depend strongly on $g(\alpha)$  for large values of 
  $\theta$. For these images, the relative errors are 
 large. For $\gamma\ge 7.0$, ${\hat{\cal Z}}(\theta)$ is such that 
 $\Delta(\theta)>0.1$ for $\theta\simr\theta_{\epsilon}$ and
 $|\delta{\hat{\cal Z}}(\theta)|/{\hat{\cal Z}}(\theta)>1.0$.
 Contrastingly,  the other images ${\cal Z}^{(\alpha)}(\theta)$  [$\gamma\le 6.0$
 and $m_{L/L_0}(\theta)$] do not vary over  the range of integration in
 Eq. (\ref{eqn:finalZ}), and these ${\hat{\cal Z}}(\theta)$ are
 independent of $g(\alpha)$, with $\Delta(\theta)<0.1$ over the entire range of $\theta$.
   The relative errors  $|\delta{\hat{\cal
 Z}}(\theta)|/{\hat{\cal Z}}(\theta)$ for these cases  [$\gamma\le 6.0$
 and $m_{L/L_0}(\theta)$]   increase with 
 $\theta$ but do not exceed 1.0. This indicates that the uncertainty
 reflects the prior information dependence of the most probable image. 
\vspace*{3mm}\par
\begin{figure}[!ht]
 \begin{center}
 \includegraphics[height=70mm]{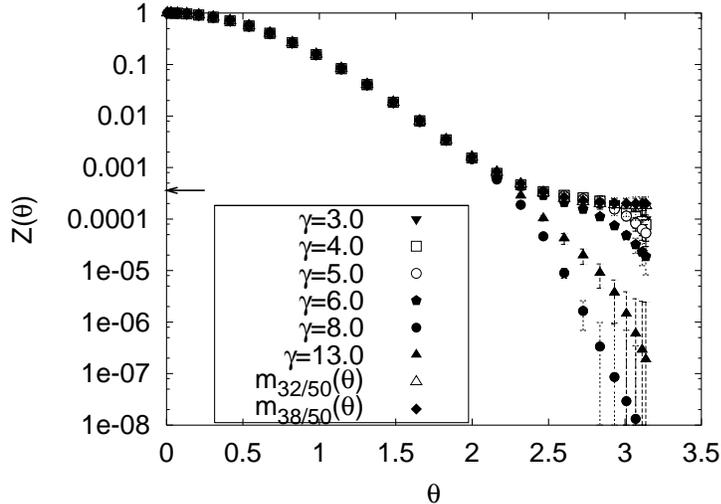}
\end{center}
\caption{The most probable images ${\hat{\cal Z}}(\theta)$ for various 
 $m(\theta)$. The arrow indicates the value of $\epsilon\hspace{.1cm}(=3.610\times 10^{-4})$.} 
\label{fig:finalZ}
\end{figure}
 Summarizing the above analysis, we present the results for  ${\hat{\cal
 Z}}(\theta)$ in Fig. \ref{fig:finalZ} for various $m(\theta)$. 
 All the results obtained using the MEM behave smoothly over the entire range of 
   $\theta$. For $\theta\siml 2.3$, these eight images fall on
 the same curve,  and the MEM reproduces  images
 consistent with the FTM. 
 For $\theta\simr 2.5$,  by contrast,  the $m(\theta)$  dependence of
 ${\hat{\cal Z}}(\theta)$ is clearly seen. In
 the large $\theta$, ${\hat{\cal Z}}(\theta)$ for $\gamma=8.0$
 and 13.0, which possess  large errors, decrease over several
 orders as $\theta$ increases, while the other images with small
 errors gradually decrease as $\theta$ increases. Each of these  images  obtained in Fig. \ref{fig:finalZ} 
 could be  a candidate for the true image. 
 With the observation  that
 ${\hat{\cal Z}}(\theta)$ depends strongly on $m(\theta)$ in the region
 where the values of ${\hat{\cal Z}}(\theta)$ are smaller than that 
 of $\epsilon$, the $m(\theta)$ dependence of ${\hat{\cal Z}}(\theta)$
 reflects the flattening phenomenon in the FTM. 
\setcounter{equation}{0}
\section{Conclusions and discussion}
\label{sec:Sum}
 In this paper, we have applied the MEM to the MC data of the CP$^3$
 model. 
 We have studied how the flattening phenomenon is observed with  the MEM.
   For this purpose, two types of data were  used, that  for $L=38$,  in 
 which no flattening is observed,  and that  for $L=50$,  in
 which flattening is reproduced through the Fourier transform.
\par
 The results we  obtained in the present study are the
 following. \\  
%
\begin{enumerate}
 \item In the case without flattening, the MEM yielded images  ${\hat{\cal
       Z}}(\theta)$ that are almost independent of $m(\theta)$ and
       $g(\alpha)$. The most probable images 
       ${\hat{\cal Z}}(\theta)$ are in agreement with the result of the
       FTM within the errors (see Fig. \ref{fig:Zalpha_L38} and Table
       \ref{table:Z_L38}).  
 \item In the case with flattening, 
       we have systematically checked  (i)
       the statistical fluctuations  of ${\hat{\cal Z}}(\theta)$, (ii)
       the $g(\alpha)$ dependence of ${\hat{\cal Z}}(\theta)$ and (iii)
       the relative error of ${\hat{\cal Z}}(\theta)$. We found that  the
       statistical fluctuations  of ${\hat{\cal Z}}(\theta)$ become 
       smaller as  the number of measurements increases except  near
       $\theta=\pi$. 
      We also found that  ${\hat{\cal Z}}(\theta)$ with large errors
       depends strongly on $g(\alpha)$ in the region of large $\theta$,
        where the $g(\alpha)$  dependence of ${\hat{\cal Z}}(\theta)$
       was  estimated using the quantity  $\Delta(\theta)$. 
       Our results are summarized in Fig. \ref{fig:finalZ}. 
       All the results obtained using the MEM behave smoothly over the entire range of 
   $\theta$. In   the region where the value
       of ${\hat{\cal Z}}(\theta)$ is larger than or nearly equal to that of $\epsilon$, 
       (specifically  $\theta\siml 2.3$), final  images fall on the same curve,  and the MEM        
       reproduces  images consistent with the FTM.   Contrastingly, in   the region where the value  of ${\hat{\cal Z}}(\theta)$ is smaller than  that of $\epsilon$, ${\hat{\cal Z}}(\theta)$ depends strongly on $m(\theta)$.   This  $m(\theta)$ dependence of ${\hat{\cal Z}}(\theta)$
 reflects the flattening phenomenon.  Each of these  images  obtained in Fig. \ref{fig:finalZ} 
 could be  a candidate for the true image. 
\end{enumerate}
\vspace*{2mm}\par
 In the present study, $\epsilon$ given by Eq. (\ref{eqn:epsilon}) turns
 out to be an approximate 
 indicator of  the influence of the error in $P(Q)$ to ${\cal
 Z}(\theta)$  in both the FTM and MEM cases. As seen in (ii) presented in 
 \S\ref{subsub:flattening}, ${\hat{\cal Z}}(\theta)$ starts to exhibit 
 the $g(\alpha)$ dependence at $\theta=\theta_{\epsilon}$ for
 $m_{\rm G}(\theta)$ with $\gamma\ge 7.0$, where $\theta_{\epsilon}$ is
 defined by ${\hat{\cal Z}}(\theta_{\epsilon})\simeq \epsilon$, and 
 $\Delta(\theta_{\epsilon})\simeq 0.1$ holds. The other
 default models investigated here satisfy $\Delta(\theta)<0.1$ for all
 $\theta$. For these, the $g(\alpha)$ dependence is very weak,   even if 
 ${\hat{\cal Z}}(\theta)<\epsilon$. It is worthwhile to study 
 the reason for this.
\par
 The magnitude of the relative error depends on $m(\theta)$ for  large
 $\theta$ (see Fig. \ref{fig:relativeerror}). At
 $\theta=\theta_{\epsilon}$, the FTM gives $|\delta{\cal Z}_{\rm
 Four}(\theta_{\epsilon})|/{\cal Z}_{\rm Four}(\theta_{\epsilon})\simeq 
 0.3$, while $|\delta{\hat{\cal
 Z}}(\theta_{\epsilon})|/{\hat{\cal Z}}(\theta_{\epsilon})\simeq 0.1$
 holds in some cases in the MEM; ${\hat{\cal Z}}(\theta)$ for
 $\gamma=3.0$, 4.0, 5.0, 6.0 and $m_{38/50}(\theta)$. Although
 $\delta{\hat{\cal Z}}(\theta)$ is the uncertainty in  the 
 image, it is necessary to elucidate the different manners in which  the error in
 $P(Q)$ affects ${\hat{\cal Z}}(\theta)$ and ${\cal Z}_{\rm
 Four}(\theta)$. When ${\hat{\cal Z}}(\theta)$ depends strongly on
 $m(\theta)$, each ${\hat{\cal Z}}(\theta)$ could be a candidate for the
 true image. If we had proper knowledge about $m(\theta)$ as prior
 information, we could identify  the true image in a probabilistic sense.
  Such  analysis may allow us  to
 clarify the relationship between the default model dependence and the 
 systematic error.
\par
 The MEM provides a  probabilistic 
 point of view in the study of theories with the sign problem. The
 canonical approach~\cite{rf:KdeF}  in the study of lattice field theory with  a  finite
 density exhibits a formal correspondence to lattice field theory with
 the $\theta$ term.  Noting this correspondence, it 
  may be worthwhile to study lattice QCD with a  finite density in terms
 of the MEM from a  probabilistic point of view.  
 
\section*{Acknowledgements}
 The authors thank  R. Burkhalter for providing his FORTRAN code for the
 CP$^{N-1}$ model with a  fixed point action. We were informed about a work concerning  the canonical  approach  in   finite density QCD, by  Prof. Ph. de Forcrand.  
 We thank him for calling our attention to the relevant  work  by him and his collaborator. 
 This work is supported in
 part by Grants-in-Aid for Scientific Research (C)(2) for  the Japan
 Society for the Promotion of Science (No. 15540249) and  the Ministry
 of Education,  Culture, Sports,   Science and Technology (Nos.  13135213 and
 13135217). Numerical calculations were  performed   at
 Computer and Network Center, Saga University.

%
\appendix
     
\end{document}